\documentclass{aa}

\usepackage{graphicx}
\usepackage[section]{placeins} 
\usepackage{multirow}

\usepackage[varg]{txfonts}  

\usepackage[english]{babel}
\usepackage[babel=true]{csquotes}
\usepackage{amssymb}   

\usepackage{stfloats}  

\newcommand{\xmm}{{\it XMM~\/}}
\def\ergsec{{\rm erg\;s^{-1}}}
\def\kms{{\rm km\;s^{-1}}}
\def\kev{{\rm keV}}
\def\cmsq{{\rm cm^{-2}}}
\def\Msun{{\rm M_{\odot}}}
\def\AAb{{\rm \AA}}
\def\ergcmsA{{\rm erg\;cm^{-2}\;s^{-1}\;\AA^{-1}}}

\begin{document}
	\title{A new sample of X-ray selected narrow emission-line galaxies.}
		\subtitle{II. Looking for True Seyfert 2.}	
	\author{E. Pons
		\and M. G. Watson}		
	\institute{Department of Physics \& Astronomy, University of Leicester, Leicester, LE1 7RH, UK\\}
	\date{Received XXX / Accepted XXX}
	\abstract{A sample of X-ray and optically selected narrow emission-line galaxies (769 sources) from the 3XMM catalogue cross-correlated with SDSS (DR9) catalogue has been studied. Narrow-emission line active galactic nuclei (AGN;  type-2) have been selected on the basis of their emission line ratios and/or X-ray luminosity. We have looked for X-ray unobscured type-2 AGN. As X-ray spectra were not available for our whole sample, we have checked the reliability of using the X-ray hardness ratio (HR) as a probe of the level of obscuration and we found a very good agreement with full spectral fitting results, with only 2\% of the  sources with apparently unobscured HR turning out to have an obscured spectrum. Despite the fact that type-2 AGN are supposed to be absorbed based on the Unified Model, about 60\% of them show no sign or very low level of X-ray obscuration. After subtraction of contaminants to the sample, that is Narrow-Line Seyfert 1 and Compton-thick AGN, the fraction of unobscured Sy2 drops to 47\%. For these sources, we were able to rule out dust reddening and variability for most of them as an explanation of the absence of optical broad emission-lines. The main explanations remaining are the dilution of weak/very broad emission-lines by the host galaxy and the intrinsic absence of the broad-line region (BLR) due to low accretion rates (i.e. True Sy2). \\
However, the number of True Sy2 strongly depends on the method used to verify the intrinsic lack of broad lines. Indeed using the optical continuum luminosity to predict the BLR properties gives a much larger fraction of True Sy2 (about 90\% of the unobscured Sy2 sample) than the use of the X-ray 2 $\kev$ luminosity (about 20\%). Nevertheless the number of AGN we securely detected as True Sy2 is at least three times larger that the previously confirmed number of True Sy2.}
\maketitle
\section{Introduction} 
The observational classes of active galactic nuclei (AGN) can be broadly divided in two categories: those which only show narrow emission lines in their spectra (i.e. type-2) and those which have both broad and narrow emission lines in their spectra (i.e. type-1). The current Unified Model (UM) for AGN has been very successful in explaining these two different types of AGN: the classification depends on the viewing angle with respect to the obscuring torus surrounding the central region of the AGN \citep{Antonucci93, Urry95}. When the central supermassive black hole (SMBH) is seen directly, both the broad-line (BLR) and the narrow-line regions (NLR) are observed (type-1 AGN), while when the line of sight is intercepted by the torus only the NLR is visible (type-2 AGN). 
	
This model is supported by the detection of polarised broad emission lines for about $50\%$ of type-2 AGN \citep{Antonucci85, Tran01} and the detection of a high X-ray column density in most of them. This supports the idea that type-1 and type-2 AGN are intrinsically the same but viewed at different angles and that the absence of broad emission lines (BELs) in type-2 AGN is due to the presence of a toroidal structure (i.e the torus) covering the central engine and the BLR.

However, several inconsistencies with the AGN UM have emerged within the last decade. There is observational evidence of objects showing opposite X-ray and optical classification. First, some optically broad emission-line (BEL) AGN show significant absorption in the X-ray band \citep{Maiolino01, Brusa03, Merloni14}. A possible explanation is that the dust to gas ratio in these objects is lower than the Galactic value due to the presence of gas within the dust sublimation radius \citep{Maiolino01}. Thus such an obscuring material primarily made of gas and with very little dust will be opaque in the soft X-rays but transparent at optical wavelengths permitting the observation of BEL. \\
On the contrary, unobscured narrow-line AGN have been reported observationally \citep{Tran01, Tran03, PanessaBassani02, Hawkins04, BrightmanNandra08, Shi10, Tran11, Huang11, Bianchi12, Merloni14} and discussed theoretically \citep{Nicastro00, Laor03, Elitzur06, ElitzurHo09}. Various explanations have been proposed to explain to the absence of observed BEL:
\begin{enumerate}
\item Variability of the level of obscuration highlighted by the non-simultaneous X-ray and optical observations,
\item Different obscuration in the X-rays and optical; i.e high dust-to-gas ratio compared with the Galactic value,
\item Dilution of weak or very broad emission-lines by the host galaxy light,
\item Intrinsic absence of the BLR at low luminosities or low accretion rates.
\end{enumerate}

The fraction of unobscured narrow-line AGN for the XMM-COSMOS field has been recently estimated as being around 30\% by \citet{Merloni14}. This is in agreement with the previous study of \citet{PanessaBassani02} who estimated that the fraction of unobscured Seyfert 2 (Sy2) varies between 10 to 30\%. These results are consistent with what was found by \citet{Mainieri02} (30\%) and \citet{Caccianiga04} (12\%). However, \citet{Risaliti99} found an even smaller fraction of only 4\%. On the contrary, \citet{Marinucci12} claimed that the fraction of unobscured Sy2 in their sample in about 40\% and \citet{Page06} and \citet{Garcet07} reported an even higher fraction of about $66\%-68\%$.\\
So the fraction of unobscured Sy2 may represent a significant part of the AGN population. The mismatch between X-ray and optical properties may also lead to the questioning the universality of the UM and seems to suggest that orientation may not be the only parameter that distinguishes type-1 and type-2 AGN.

In this paper, we have looked at the fraction of unobscured Sy2 within a new 3XMM-SDSS cross-matched catalogue. Section 2 describes the narrow-line (NL) AGN sample, the selection of X-ray unobscured Sy2 AGN is explained in Section 3. Possible explanations for the absence of observed BEL are investigated in Section 4.


\section{The optical and X-ray AGN sample}
	\subsection{The narrow emission-line galaxies sample}

To create our AGN sample, we have cross-matched the 3XMM-DR4 catalogue \citep{Rosen16} with the Sloan digitial sky survey (SDSS) DR9 spectroscopic release \citep[see also][]{PW14}.\\
Firstly, from the SDSS catalogue we have selected objects which were spectroscopically identified as GALAXY or QSO and which are also labelled as \enquote{science primary} objects (i.e. those that have the best available spectra). We have restricted our selection to objects having spectral fitting performed by the MPA-JHU group (called the galspec measurements and described in \citet{Brinchmann04, Kauff03a, Tremonti04}).This catalogue has the advantage of including emission line fluxes corrected for Galactic reddening. In addition, through fitting of the observed spectrum with galaxy spectra models, the derived products provide estimates of emission line strengths after subtraction of stellar absorption line components. This is important because the SDSS aperture of the spectroscopic fibre is 3" and thus the spectra include not only the nucleus but have also host galaxy contribution. Moreover, using stellar population models, the derived products include a variety of galaxy parameters (e.g. stellar masses, stellar and gas kinematics, velocity dispersions \ldots).

Secondly, the cross-matched X-ray/optical sample was obtained by cross-correlating the SDSS galspec sub-sample with the 3XMM catalogue. We have assumed that the X-ray and optical sources correspond to the same object if the separation between the optical and X-ray positions was smaller than 10'' and for which the normalised separation (i.e. the ratio of the separation to the X-ray position error) is smaller than four \citep{Pineau11}. Because we restricted our sample to SDSS spectroscopic objects, which are significantly brighter and thus have lower sky density than  galaxies in the SDSS photometric catalogue, spurious matches become negligible even with these generous limits. Then, we required that the X-ray sources were point-like (3XMM catalogue parameter SC\_EXTENT $<$ 5; thus removing galaxy clusters) and have relatively high detection significance (catalogue parameter SC\_DET\_ML $>$ 15).\\
From this sample, narrow emission-line galaxies were selected by requiring the full-width at half maximum (FWHM) of the Balmer lines to be smaller than 1000 $\kms$ \citep[e.g.][]{Caccianiga08}; this threshold rejects all bona fide broad-line objects and reduces the contamination by conventional Narrow Line Seyfert 1 (NLS1). Narrow emission-line galaxies correspond at this stage to 1430 sources.

Finally, in order to have a trustworthy sample classification, galaxies were only considered if they had a reasonable quality optical spectrum (signal to noise ratio of the whole spectrum $S/N > 3$) and a redshift $z<0.4$ (so that all of the four lines H$\beta$, [OIII]$\lambda$5007, H$\alpha$ and [NII]$\lambda$6584 are covered). We further required that these lines were in emission with a reliable measurement; specifically the line ratios $\mathrm{[OIII]/H}\beta$ and $\mathrm{[NII]/H}\alpha$ must be three times larger than the error on the ratio. This leads to a sample of 769 narrow emission-line galaxies (NELG).

	\subsection{X-ray and optically selected AGN}

Among these narrow emission-line galaxies, AGN were selected based on the X-ray luminosity and the optical emission-line ratios. In the optical band, we considered a galaxy as an AGN if its emission-line ratios places it in the AGN region of the standard BPT diagram \citep*{BPT81}. While in the X-rays, we used the rest-frame hard X-ray (i.e. 2-10 keV) luminosity ($\mathrm{L_{HX}}$) to identify AGN. Recently, \citet{BrightmanNandra11} have found that a hard X-ray luminosity of $10^{41}\;\ergsec$ rather than the traditional $10^{42}\;\ergsec$, is an effective discriminator for AGN activity, with contamination by SF galaxies being only 3\%. So for the galaxies which are optically identified as LINERs or Composite objects by the BPT diagram, and are so unlikely to be pure SF galaxies, the X-ray luminosity threshold to select X-ray AGN can be lowered to $10^{41}\;\ergsec$.

\begin{figure}
	\centering
	\includegraphics[width=\hsize]{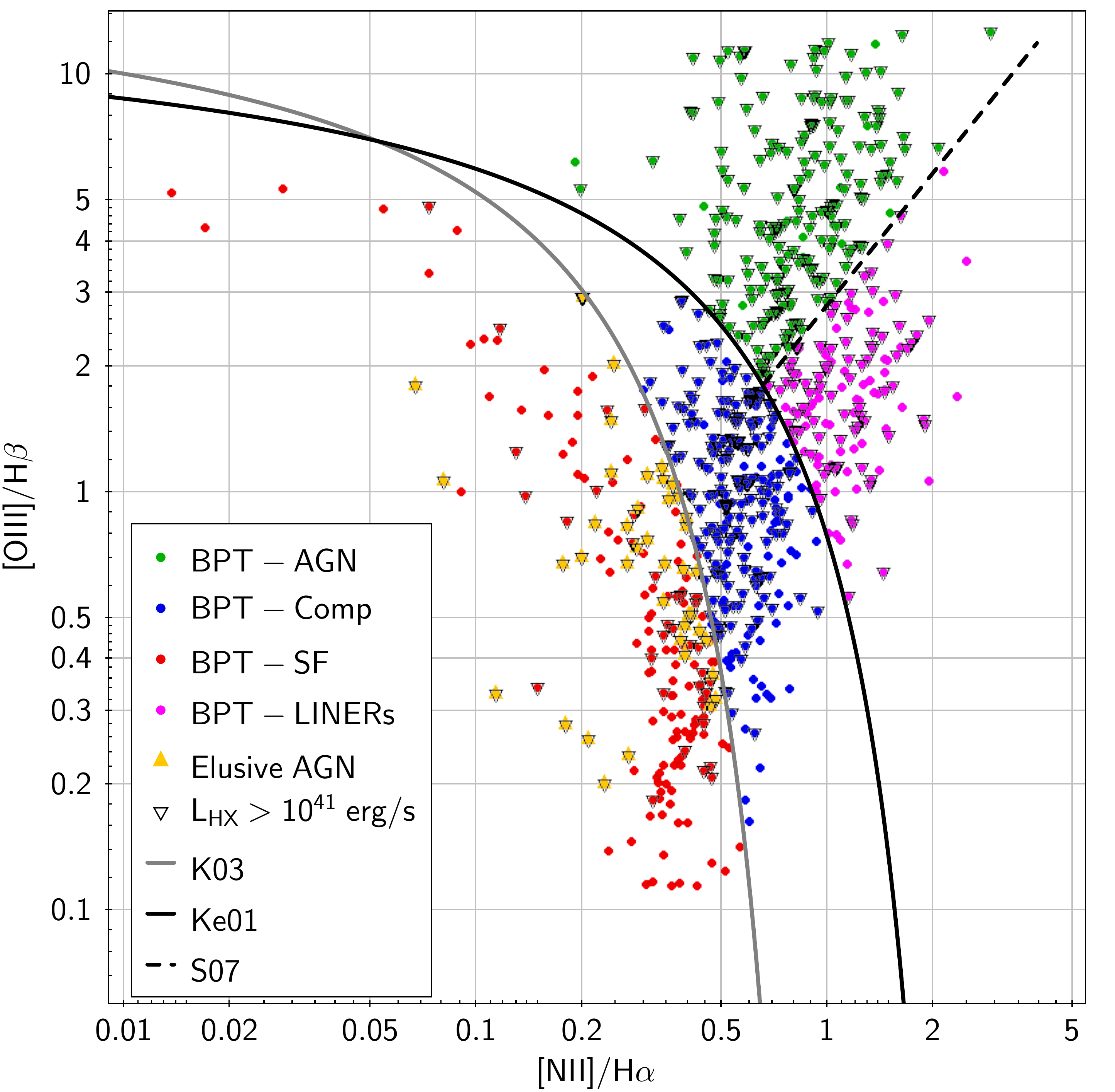}
	\caption{BPT emission line diagnostic diagram. The K03 line in grey separates the SF galaxies (red symbols) and composite objects (blue symbols) regions, while the Ke01 demarcation line in black distinguishes between AGN/LINERs and composite objects. The black dashed line divides between optically classified AGN (green symbols) and LINERs (magenta symbols). The \emph{elusive} AGN studied in \citet{PW14} are shown by yellow triangles.}
	\label{fig:BPT}
\end{figure}

In Figure \ref{fig:BPT} the BPT diagram for the NELG sample is plotted. In addition to the \citet[hereafter K03]{Kauff03b} and \citet[hereafter Ke01]{Kewley01} demarcation lines to divide between SF galaxies, composite objects and AGN, another separation line \citep[hereafter S07,][]{Schawinski07} is added to distinguish between AGN and LINERs \citep[hereafter S07,][]{Schawinski07}.\\
From the BPT diagram, the NELG sample can be divided into 202 BPT-AGN (LINERs excluded, 27\%), 138 BPT-LINERs (18\%), 239 BPT-Composite objects (31\%) and 190 BPT-SF (24\%).

Our AGN sample is composed of optically selected AGN from the BPT diagram (202 sources) plus optical BPT-Composite objects or BPT-LINERs with $L_{HX}>10^{41}\;\ergsec$ (224 sources) in addition to optical BPT-SF with $L_{HX}>10^{42}\;\ergsec$ (41 sources, corresponding to the \emph{elusive} AGN studied by \citet{PW14}). As we are looking for a sample of NL AGN (i.e. type-2 AGN), we removed the 24 \emph{elusive} sources identified as NLS1 by \citet{PW14}. This gives a \emph{NL AGN} sample of 443 sources selected based on optical and X-ray criteria.


\section{X-ray Unobscured Narrow-Line Seyfert 2}
	
	\subsection{Selection of X-ray unobscured sources}
As the X-ray spectra are not always available, the X-ray hardness ratio (HR) is often used to estimate the level of obscuration in AGN \citep{Rovilos11}. This is due to the fact that the HR variations are dominated by obscuration and that the spectra of an absorbed AGN are harder than unabsorbed one \citep{Wang04, Szokoly04}.

\begin{figure}
	\centering
	\includegraphics[width=\hsize]{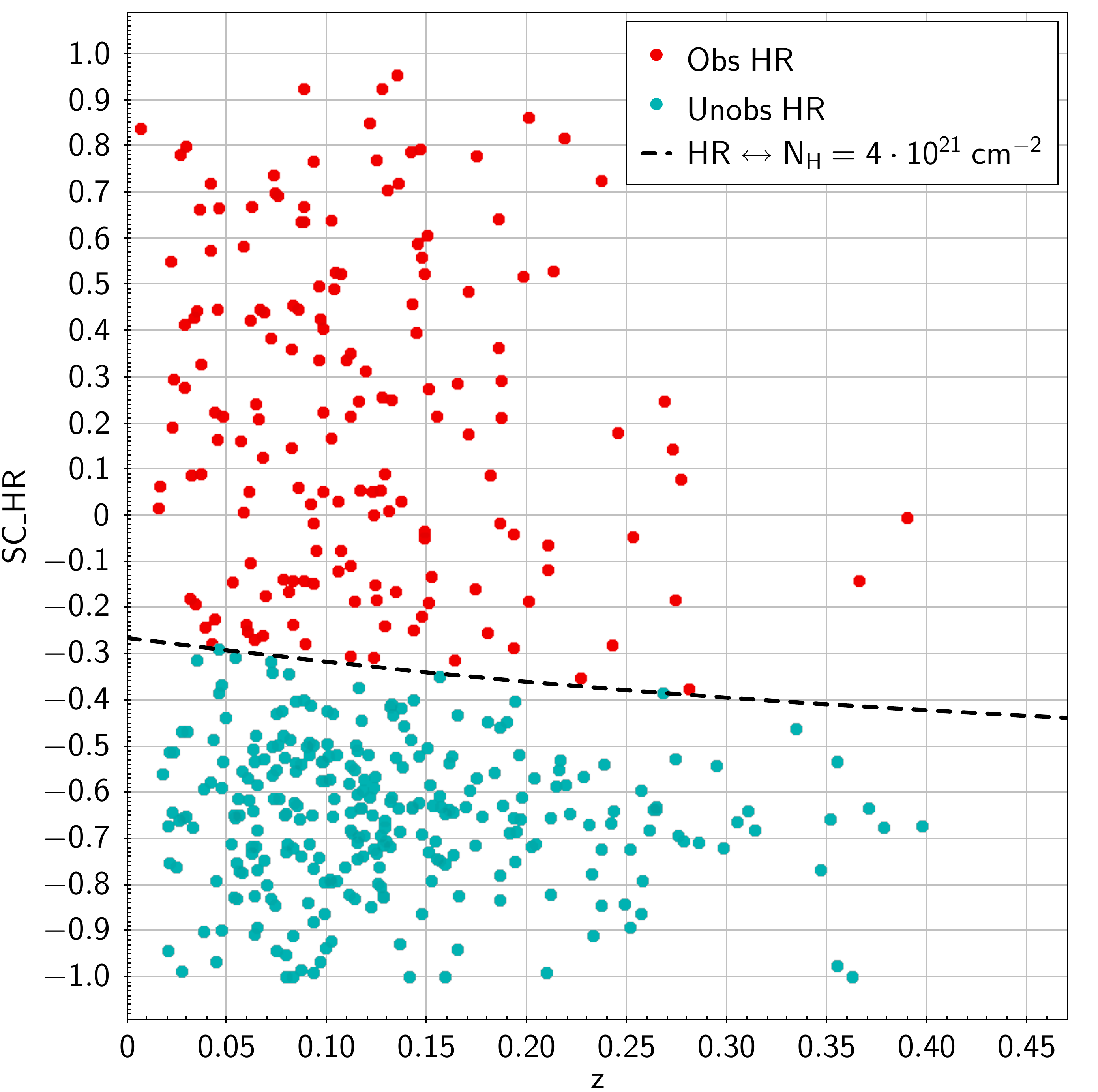}
	\caption{XMM HR vs redshift for the sources of the type-2 AGN sample. The black dashed line corresponds to the simulated HR for an intrinsic absorption of $4\times 10^{21}\;\cmsq$. Only 37\% of the sample appear to be obscured (red symbols) while 63\% have an unobscured HR (cyan symbol).}
	\label{fig:unobsSample_HRz}
\end{figure}

In the first instance, we checked the accuracy of a selection of unobscured sources based on the HR compared to the column density ($N_H$) from the spectral fit. \\
In \citet{PW14}, we used HR2 (ratio between the $(0.5-1.0)\;\kev$ and the $(1.0-2.0)\;\kev$ bands) and HR3 (ratio between the $(1.0-2.0)\;\kev$ and the $(2.0-4.5)\;\kev$ bands) hardness ratios from \xmm to estimate the level of obscuration. However in order to have a more accurate distinction between X-ray obscured and X-ray unobscured AGN and because the classification depends on the redshift \citep[see][]{Wang04, Szokoly04}, we have computed a new HR between the $(0.5-2.0)\;\kev$ and the $(2.0-4.5)\;\kev$ energy bands. Using this approach, with only one HR instead of two, it is easier to take into account the redshift evolution. \\
In order to relate the \xmm HR to the hydrogen column density $N_H$, we have used the PIMMS simulator \footnote{http://heasarc.gsfc.nasa.gov/docs/software/tools/pimms.html}. The X-ray spectra for a source with a redshift varying from 0 to 0.4 have been simulated assuming a power-law with an photon index of 1.8 modified by foreground Galactic absorption (fixed at $N_{H,Gal}\sim 1.7\times 10^{20}\;\cmsq$ which is the mean value for our sample) and by intrinsic absorption ($N_H$). The simulated HR for a given $N_H$ and redshift was then computed from the output count rates in the $(1.0-2.0)\;\kev$ and the $(2.0-4.5)\;\kev$ bands. \\
An $N_H$ of $4\times 10^{21}\;\cmsq$ was used as the separation between X-ray obscured and unobscured AGN \citep{Caccianiga07, Merloni14}. In Figure \ref{fig:unobsSample_HRz}, we plot the HR compared to the simulated HR for an intrinsic $N_H$ of $4\times 10^{21}\;\cmsq$: 279 sources (63\%) of the \emph{NL AGN} sample have an obscured HR while 163 sources are obscured.

For the 3XMM spectra, fitting has already been performed by \citet{Corral15} (the XMMFITCAT catalogue) for all sources where the net number of counts after background subtraction per detector is greater than 50 in the total band. In the case where the total number of counts is smaller than 500, they only fit simple models (i.e. an absorbed power-law, an absorbed thermal and an absorbed black-body model) in the full, soft and hard bands. Otherwise, two component models (an absorbed thermal plus power-law, an absorbed double power-law and and absorbed black-body plus power-law model) were also fitted in the total band. \\
About half of the \emph{NL AGN} sample (212 sources, 48\%) are in the XMMFITCAT catalogue and 67\% (143 sources) of these are classified as unobscured based on the HR. We used the estimated $N_H$ from the best-fit model to compare with the HR classification and found that  only 2\% (three sources misclassified) of the unobscured HR sources eventually have an obscured spectrum (i.e. $N_H>4\times 10^{21}\;\cmsq$). Only 4\% (three sources misclassified) of the obscured HR sources have $N_H<4\times 10^{21}\;\cmsq$ in the fits.

Thus the HR provides a quite reliable classification as X-ray obscured or unobscured for our sample. From 279 sources selected as unobscured based on the HR (see Figure \ref{fig:unobsSample_HRz}), we remove the three sources identified as obscured by the X-ray spectral fitting and add the three other misclassified as obscured by the HR but which present no sign of obscuration in their X-ray spectra. \\
Thus, 279 sources (63\%) of the \emph{NL AGN} sample show no sign of X-ray obscuration and form the X-ray unobscured NL AGN sample.

	\subsection{Sources misclassified as unobscured or as NL AGN}
In order to have a reliable estimate of the fraction of truly unobscured Sy2, contaminants to the sample (i.e. type 1 AGN and Compton-thick sources) need to be removed.

	\subsubsection{Contamination by NLS1}

As shown in \citet{PW14, Castello12}, narrow-line Seyfert 1 can contaminate the NL sample as their line widths can extend down to $400\;\kms$, below the normally adopted threshold ($1000\;\kms$) separating narrow and broad line AGN. However, as their broad component originates from the BLR, their Balmer line widths are expected to be larger than their forbidden line widths which are from the NLR. Being type-1 AGN, NLS1 are also expected to show broad wings for the Balmer emission lines which can be confirmed by visual inspection of the optical spectra. In addition this allows the estimated width of the line to be checked. Indeed, five sources with weak broad wings seem to have their Balmer line widths underestimated by the galspec fit (so have fitted $FWHM_{Balmer}\sim FWHM_{forbidden}$) and are likely to be NLS1. 

From the 57 sources with $FWHM_{Balmer}>1.15 \times FWHM_{forbidden}$ (but also with $FWHM_{Balmer}>400\;\kms$), 35 sources have clear weak broad wings and four more sources have too noisy spectra to confirm the presence of wings but the line fits are still reliable, so all these sources can be considered as NLS1. \\
There are also some AGN in our sample (12 sources) which are fitted with $FWHM_{Balmer}>1.15 \times FWHM_{forbidden}$ but which have a strong galaxy contribution and very weak emission-lines or very noisy spectra.  Their Balmer line widths seem to be overestimated, thus these sources are unlikely to be NLS1. The six remaining sources (with $FWHM_{Balmer}>1.15 \times FWHM_{forbidden}$) have no sign of a broad wings and have in fact consistent line widths when the errors are considered. So we do not class them as NLS1.

\begin{figure}
	\centering
	\includegraphics[width=\hsize]{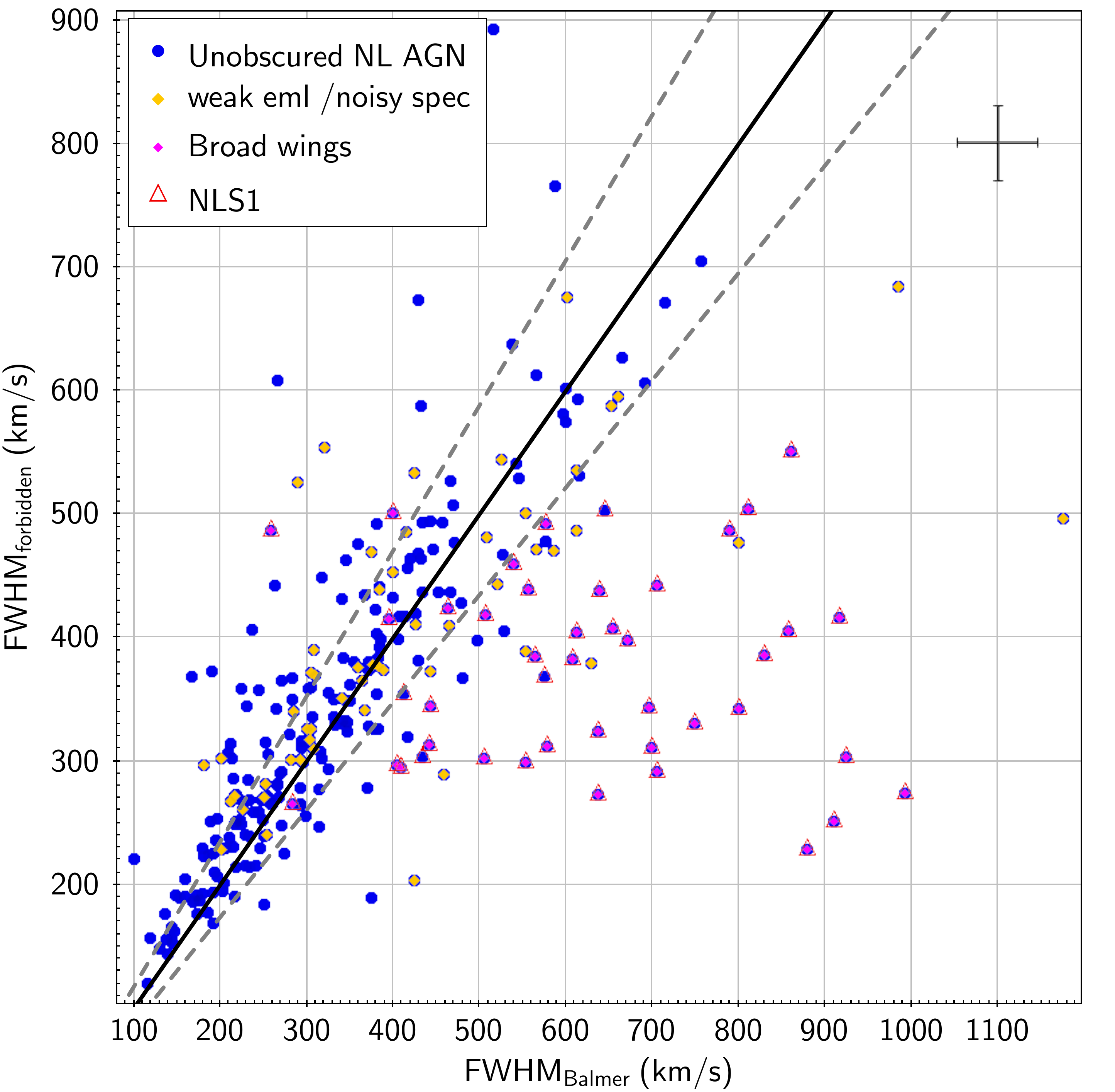}
	\caption{Balmer vs. forbidden line widths for the unobscured NL AGN. The black line is the equality line, while the two dashed grey lines correspond to an error of 15\%. The sources with noisy spectra and/or weak emission-lines and thus having an unreliable fit are shown by yellow symbols. The sources with observed weak broad wings for the Balmer lines are labelled in magenta. Finally the AGN which we classified as NLS1 (44 sources) are shown by open red triangles.}
	\label{fig:unobsSample_NLS1}
\end{figure}

In summary, we found 44 sources (16\% of the initial unobscured NL AGN sample) which are very likely to be NLS1, including 39 sources with $FWHM_{Balmer}>1.15 \times FWHM_{forbidden}$ and five sources with underestimated fitted Balmer lines; most of them showing weak broad wings in their optical spectra (see Figure \ref{fig:unobsSample_NLS1}).

In type-1 AGN and so in NLS1, permitted blended FeII emission \citep{Pogge00} can also be observed from the blue and red sides of the H$\beta$ and [OIII]$\lambda$5007 lines respectively. However in order to detect it, it requires high signal-to-noise spectra (S/N$\gtrsim 20$) which is the case for only six of our 44 selected NLS1. This discriminator is thus of very limited use for our sample.

	\subsubsection{Contamination by CT AGN}
In CT AGN ($N_H>5\times 10^{24}\;\cmsq$), the direct nuclear photons below 10 keV are completely suppressed and the X-ray spectrum is dominated by a reflected component leading to a misclassification of the source as unobscured as the absorbed power-law is no longer visible. As with the X-ray spectra, the HR also can give an inaccurate classification as unobscured.

\begin{figure}
	\centering
	\includegraphics[width=\hsize]{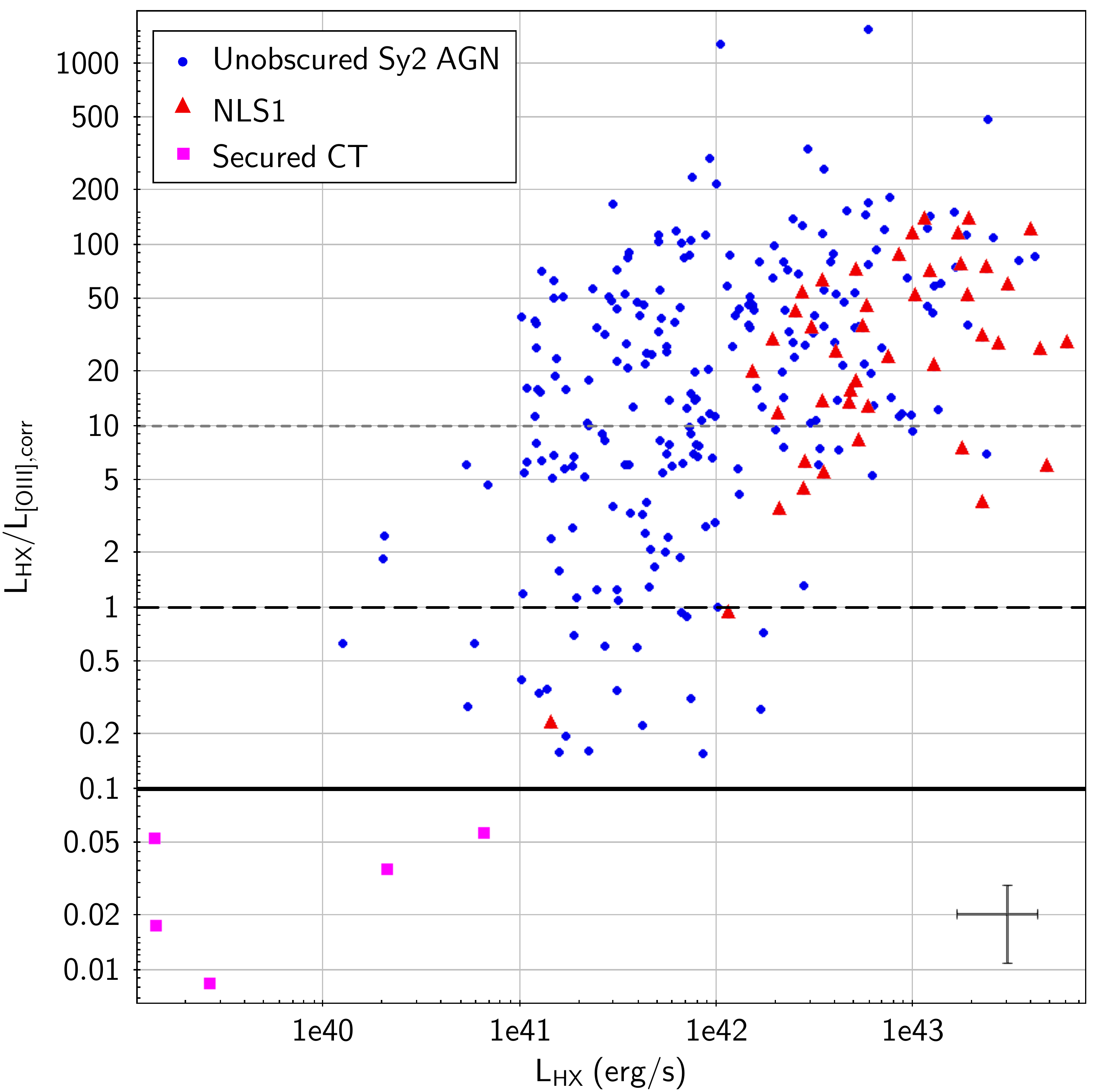}
	\caption{Comparison between the hard X-ray and the de-reddened [OIII] luminosities. Only 5 sources (magenta squares) are securely classified as CT AGN ($\mathrm{L_{HX}/L_{[OIII], corr}}<0.1$, black line), while 30\% of the sample still fall in a region where CT AGN can be found ($0.1<\mathrm{L_{HX}/L_{[OIII], corr}}<1$, black dashed line and $1<\mathrm{L_{HX}/L_{[OIII], corr}}<10$, grey dashed line.) NLS1 are represented by red triangles.}
	\label{fig:unobsSample_CT_OIII}
\end{figure}

In order to identify CT AGN, two indicators of the AGN luminosity need to be compared, one of which is affected by the obscuring material while the other is unchanged. As a first example, in the presence of a high level of absorption, the X-ray luminosity will be depressed by an amount depending on $N_H$ while the [OIII] luminosity originating from the NLR will be unaffected. Compton-thick sources are thus expected to have a $\mathrm{L_{HX}/L_{[OIII], corr}}$ ratio smaller than 0.1 \citep{Bassani99}. As we can see in Figure \ref{fig:unobsSample_CT_OIII}, only five sources of the unobscured sample are found with such a low ratio. \\
However, recent work has shown that CT objects may have a higher ratio than the standard 0.1 value used. \citet{BrightmanNandra11} have argued that the $\mathrm{L_{HX}/L_{[OIII], corr}}$ may be an unreliable parameter to detect CT AGN as they can have a ratio as high as ten. In addition, \citet{Goulding11} have found that $0.1<\mathrm{L_{HX}/L_{[OIII], corr}}<1$ may be an ambiguous region, as CT objects may often be found with this range of ratio.\\
In our \emph{unobscured NL AGN} sample, in addition to the five secure CT with $\mathrm{L_{HX}/L_{[OIII], corr}}<0.1$, 21 sources have $0.1<\mathrm{L_{HX}/L_{[OIII], corr}}<1$ and 64 more have $1<\mathrm{L_{HX}/L_{[OIII], corr}}<10$.
\\

\begin{figure}
	\centering
	\includegraphics[width=\hsize]{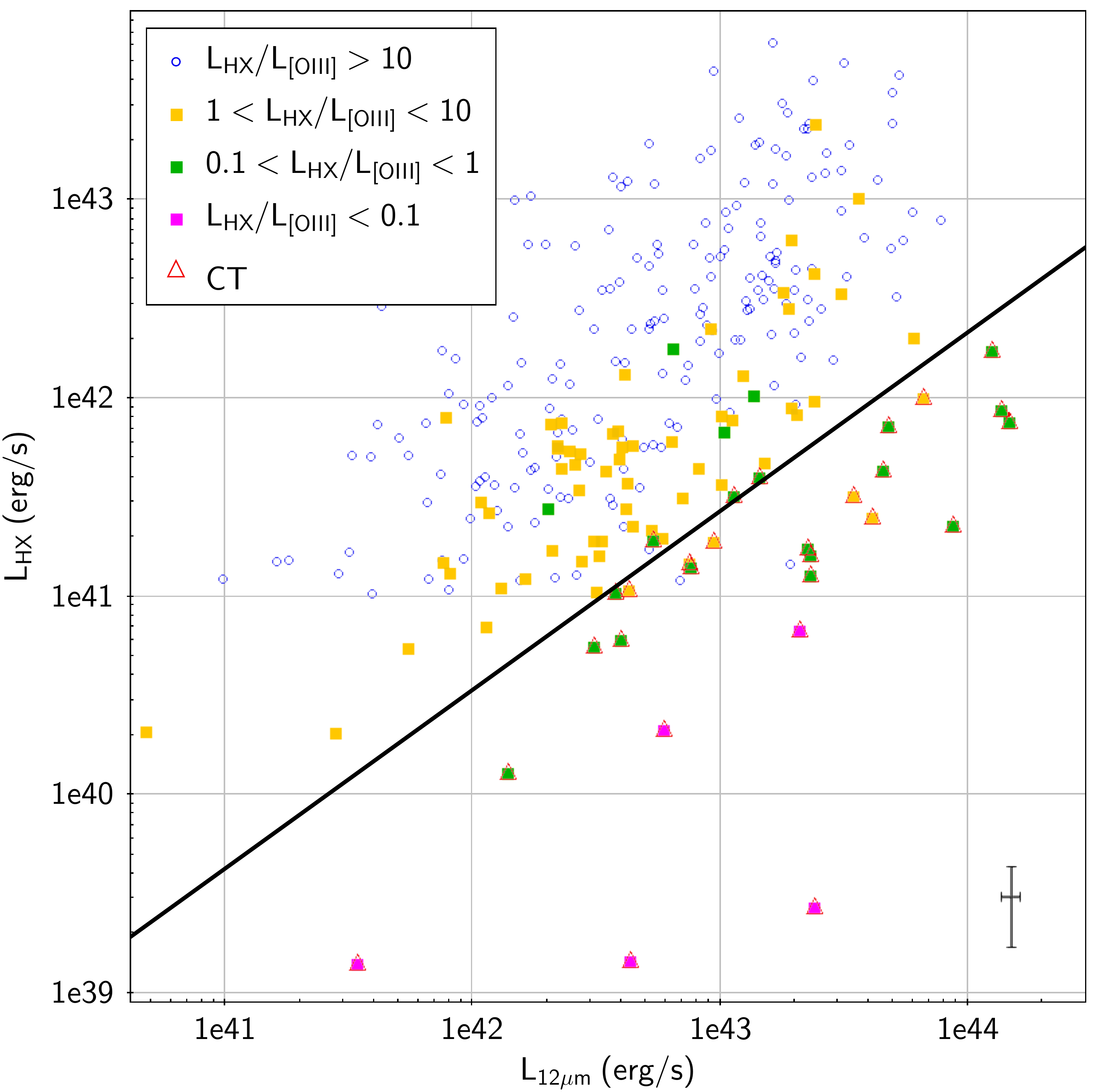}
	\caption{Hard X-ray vs. 12$\mu m$ luminosities. 28 unobscured Sy2 AGN (red triangles) are found to be below (or close to) the CT demarcation line (black line; corresponds to the AGN relation from \citet{Gandhi09} scaled down by a factor 25) and are thus classified as CT AGN.}
	\label{fig:unobsSample_CT_MIR}
\end{figure}

On the other hand, the mid-IR luminosity, combined with the hard X-ray luminosity, has been considered as a better indicator to test the presence of heavy absorption instead of the  $\mathrm{L_{HX}/L_{[OIII], corr}}$ ratio \citep{Alexander08, Goulding11, Georgantopoulos11, Rovilos14}. The MIR is associated with reprocessed emission by dust and, around 10$\mu m$, corresponds to the peak of the AGN-powered IR emission which also coincides with the minimum contribution from the host galaxy starlight \citep{Chary01, Bongiorno12}. \\
All the sources in our sample are detected by WISE, and their observed magnitudes at 3.4$\mu m$, 4.6$\mu m$, 12$\mu m$, 24$\mu m$ are thus available. In most previous work \citep{Goulding11, Georgantopoulos11} the 6$\mu m$ luminosity is used at a proxy of the AGN luminosity. However this requires extrapolation from the four WISE bands assuming a power-law SED (i.e. $f_{\nu} \propto \lambda^{a}$). As the SED constrained only by the WISE data may be more complex for some sources, we instead used the relation between the hard X-ray and the 12$\mu m$ luminosities for AGN of \citet{Gandhi09}, like \citet{Rovilos14}, to find CT AGN. The $\mathrm{L_{12\mu m}}$ luminosity can be directly inferred from the W3 WISE magnitude. No K-correction was applied but this effect should be negligible because we are dealing with low redshift sources ($z<0.4$), especially compared with the uncertainties in the SED fitting. Indeed, for a SED power-law slope of 1.5, the difference will be only of 18\% at a redshift 0.4.\\
Based on the PIMMS simulator, for a CT AGN (i.e. $N_H>5\times 10^{24}\;\cmsq$), the ratio between the observed and the intrinsic X-ray flux is $f_{HX, obs} / f_{HX, int}\sim 0.04$. So we adopted the $L_{HX}-L_{12\mu m}$ AGN relation of \citet{Gandhi09} and scaled it down by a factor 25 to have the expected limit for CT AGN.

As we can see in Figure \ref{fig:unobsSample_CT_MIR}, as expected all the sources with $\mathrm{L_{HX}/L_{[OIII], corr}}<0.1$ are in the $L_{HX}-L_{12\mu m}$ CT region. In addition, 17 (81\%) of the sources with $0.1<\mathrm{L_{HX}/L_{[OIII], corr}}<1$ and six (9\%) of those with $1<\mathrm{L_{HX}/L_{[OIII], corr}}<10$ are identified as CT AGN based on the 12$\mu m$ luminosity. These results confirm that for $0.1<\mathrm{L_{HX}/L_{[OIII], corr}}<1$, a large fraction of the sources are indeed CT AGN and that using the $\mathrm{L_{HX}/L_{[OIII]}}$ ratio only may underestimate the number of CT AGN.\\
So in the unobscured Sy2 sample (after exclusion of the NLS1), the fraction of CT objects is 12\% (i.e. 28 sources).
\\
Thus the final \emph{unobscured Sy2 sample}, after exclusion of the sources misclassified as type-2 (i.e. the 44 NLS1) or misclassified as X-ray unobscured (i.e. 28 CT AGN), is composed of 207 sources corresponding to 47\% of the initial \emph{NL AGN} sample.

	\section{Why broad emission-lines are not observed in X-ray unobscured AGN?}

Several factors may be able to explain the absence of observed broad emission-lines (BEL) in the optical spectra of truly X-ray unobscured AGN. These include: (1) non-simultaneous X-ray and optical observations, (2) different obscuration in the X-rays and optical, i.e. dust extinction, (3) dilution of weak BL by the host galaxy and (4) intrinsic absence of the BLR at low luminosity and/or low accretion rates AGN.

	\subsection{Non-simultaneous X-ray and optical observations}
	\label{sec:var}

Some AGN appear to change their optical or their X-ray classification (i.e. from narrow-line/obscured to broad-line/unobscured or vice-versa) over time as a result of obscuring clouds moving along the line of sight \citep{Tran92, Elvis04, Risaliti05, BianchiPiconcelli09}. In this case, a disagreement between the X-ray and optical classifications may be apparent if the observations in the two bands are made at different epochs.

In order to explain the reddening of the BLR, the absorbing material must be situated outside the sublimation radius of the AGN, implying variability on timescales of years to decades \citep{Bianchi12b}. The time difference between the X-ray and optical observations for the \emph{unobscured Sy2} sources in the sample range from few days to 11 years and so may be affected by variability.\\
\citet{Bianchi12} have estimated the minimum time needed for a cloud to completely cover and uncover the BLR, assuming that it is moving with a Keplerian motion around the BH (see ep. \ref{eq:vartm}). This parameter depends both on the $M_{BH}$ and the bolometric luminosity $L_{bol}$ ($L_{bol}=L_{HX}\cdot C_{HX}$) of the AGN which are derived using the relationships of \citet[see eq. \ref{eq:MBH_LKtot}]{Lasker14} and \citet[see eq. \ref{eq:Lbol}]{Marconi04} respectively.

\begin{equation}
	\centering
	\log M_{BH}=8.36+0.92\left(\log L_{K,tot}-11\right)
	\label{eq:MBH_LKtot}
\end{equation}
where $M_{BH}$ and $L_{K,tot}$ are in units of $M_{\odot}$ and $L_{\odot}$ respectively. The BH mass is inferred here from the total rest-frame K-band luminosity as this is dominated by host galaxy starlight \citep{AWK96}. We have used the K-band luminosity from 2MASS or UKIDSS in this work.

\begin{equation}
	\centering
	\log \left(\frac{L_{bol}}{L_{HX}}\right)=1.54+0.24\mathcal{L}+0.012\mathcal{L}^2-0.0015\mathcal{L}^3
	\label{eq:Lbol}
\end{equation}
where $\mathcal{L}=\log(L_{bol})-12$, and $L_{bol}$, $L_{HX}$ are in units of $L_{\odot}$.

\begin{equation}
	\centering
	t_{min}(s) = 7.9 \times 10^{7} \left(\frac{L_{bol}}{10^{43}\;\ergsec}\right)^{3/4} \left(\frac{M_{BH}}{10^{8}\;\Msun}\right)^{-1/2}
	\label{eq:vartm}
\end{equation}
\\
Based on the comparison between the observation time differences and $t_{min}$ for the multiple \emph{XMM} observations or the multiple SDSS data sets available, variability can be ruled out for 37\% of the sample. Moreover it is also very unlikely for 10\% of the sources as the multiple X-ray observations do not detect $N_H$ variability or multiple optical sets do not reveal BEL variation within the expected timescale. Only 1 source is found to have a variable HR and X-ray classification between several \emph{XMM} observations.\\
Thus the non-simultaneous X-ray and optical observations are unable to explain the absence of BEL for 48\% (100 sources) of the sample, but there are still 106 \emph{unobscured Sy2} for which  variability cannot be ruled out.

	\subsection{Different obscuration in the X-rays and optical}

It have been suggested previously \citep{PanessaBassani02, Huang11} that an extremely high dust-to-gas (D/G) ratio compared to the Galactic value may be able to explain the absence of BL. Indeed, obscuration by a dusty absorber which has a high fraction of dust associated with very little gas, will redden the BLR but produce only small effects on the X-ray properties. Within the host galaxy, large scale structures like dust lanes \citep{Rigby06} or star-forming regions may be able to hide the BLR.
\citet{Maiolino01} have found that deviation from the standard galactic gas-to-dust (G/D) ratio is very common in AGN, but they observed more gas than expected for intermediate and high luminosity ($L_{HX}>10^{42}\;\ergsec$) AGN  while the opposite scenario is needed (i.e. more dust) to redden the BLR without affecting the X-ray spectra. They only reported smaller G/D ratio (i.e. higher D/G ratio) than the Galactic value for low-luminosity AGN.

The presence of a abnormal high D/G ratio can be tested by comparing the column density inferred from the X-ray spectra with the extinction to the NLR, assuming the relation of \citet{Ward87} between the optical extinction and the Balmer decrement :
$$A_V = 6.67 \left(\log{(H\alpha/H\beta)}-\log{2.85}\right)$$

This can then be translated to a hydrogen column density assuming the Galactic relation of \citet{Guver09}:
$$A_V = 4.5\times 10^{-22} N_{H}$$

\begin{figure}
	\centering
	\includegraphics[width=\hsize]{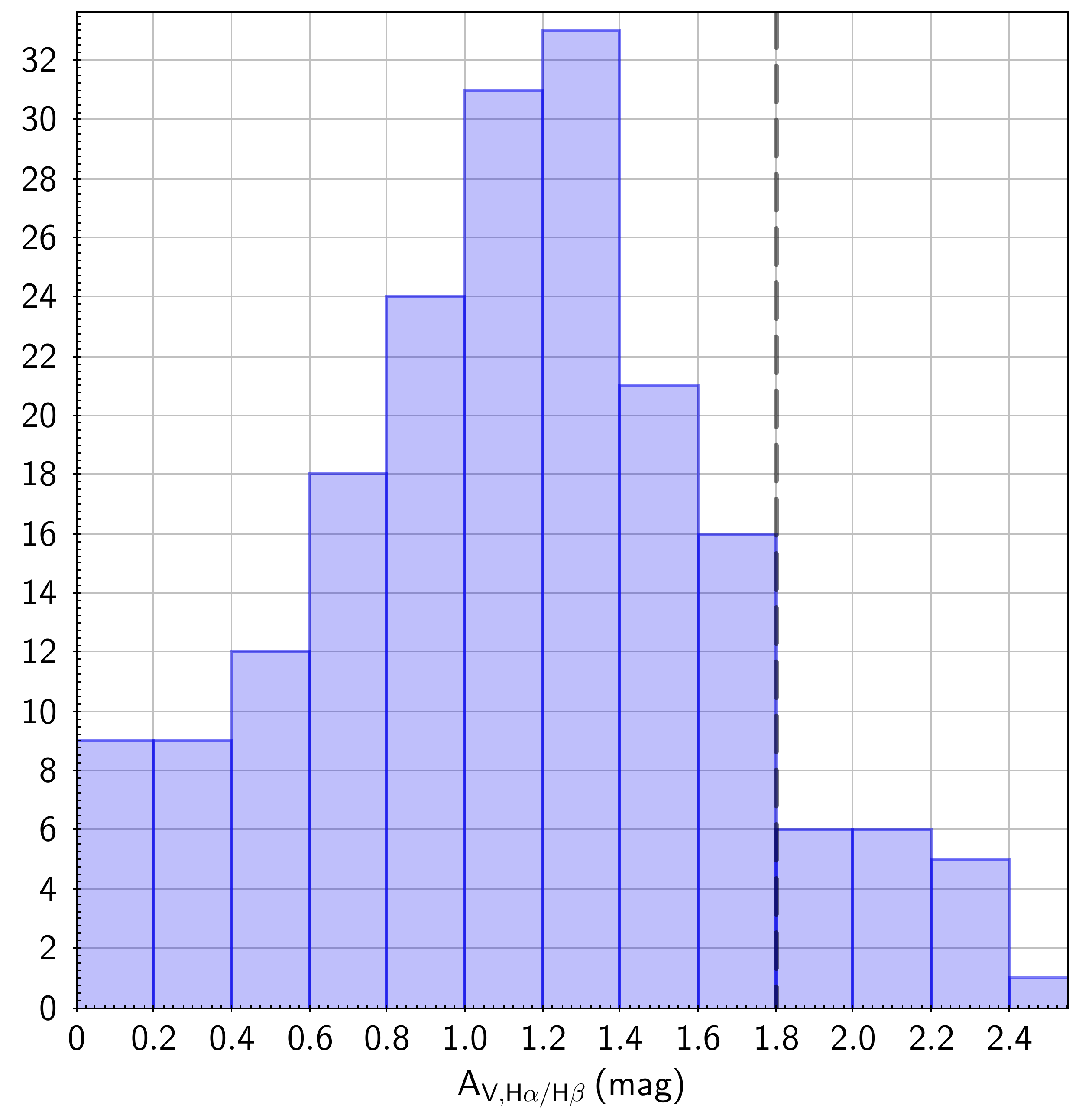}
	\caption{Distribution of the optical extinction inferred from the Balmer decrement for the \emph{unobscured Sy2}. Only 18 sources may be considered as obscured ($N_{H} > 4\times 10^{21}\;\cmsq \leftrightarrow A_V >1.8\;\mathrm{mag}$) due to a higher G/D ratio than the Galactic even if the maximum observed optical extinction is quite low $A_{V,max}\sim 2.5\mathrm{mag}$.}
	\label{fig:unobsSample_Av}
\end{figure}
	
For the \emph{unobscured Sy2}, the optical extinction is in the range  $A_V \sim 0.01-2.46\;\mathrm{mag}$ with a peak at $1.2\;\mathrm{mag}$ (see Figure \ref{fig:unobsSample_Av}), which corresponds to $N_{H,Av}\sim 2.3\times 10^{19} - 5.4\times 10^{21}\;\cmsq$. If we compare the $N_{H,Av}$ with the value from the fit for the 95 sources which are in the XMMFITCAT, we have $few<N_{H,Av}/N_{H,Fit} <40$ corresponding to a similar or higher D/G ratio than the Galactic. Thus, it seems to confirmed that unobscured type-2 AGN have a higher D/G ratio than the standard Galactic relation.\\
However, even with a higher D/G ratio, there are only 18 sources for which $N_{H,Av} > 4\times 10^{21}\;\cmsq$ (i.e. $A_V >1.8\;\mathrm{mag}$) and can be considered as \enquote{obscured}. Moreover, the maximum optical extinction is only $\sim 2.5\;\mathrm{mag}$ and so is very unlikely to hide a strong BEL, which usually requires $A_V > 10\;\mathrm{mag}$ \citep{Veilleux97}.

	\subsection{Dilution by the host galaxy}
	\label{sec:unobsDil}
In the case of weak type-1 AGN or for extremely broad lines, the BEL can be overwhelmed by the host galaxy starlight in addition to being hidden in low S/N optical spectra. Thus the dilution by the galaxy may explain why, even if the AGN is observed directly and so is not obscured by the torus, no BEL are observed \citep{Merloni14}.

\begin{figure}
	\centering
	\includegraphics[width=\hsize]{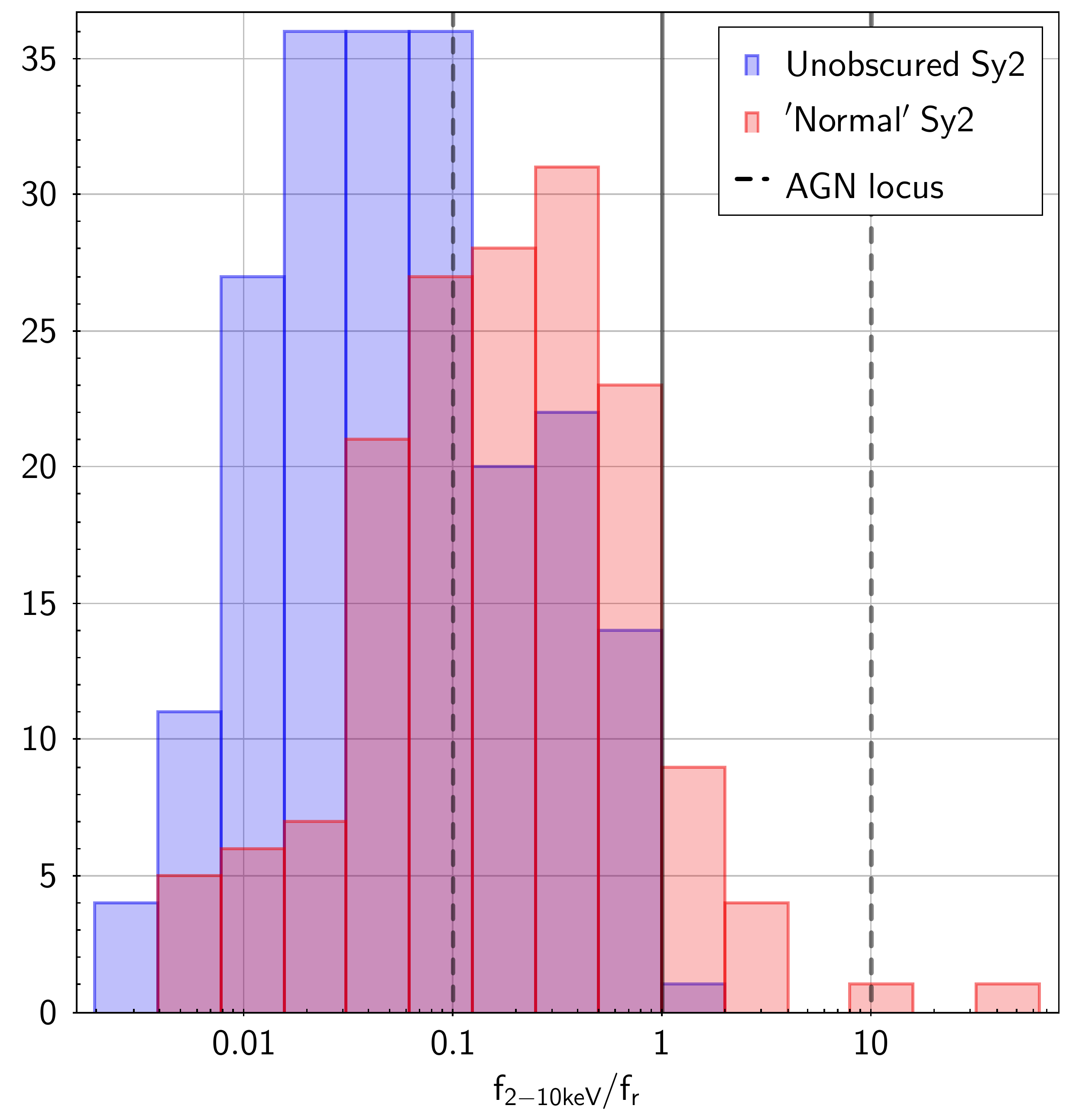}
	\caption{Distribution of the X-ray to optical flux ratio. Compared to normal obscured Sy2 (red), \emph{unobscured Sy2} (in blue) peak at a lower X/O ratio, confirming the decreasing contrast between the AGN light and the host galaxy starlight.}
	\label{fig:unobsSample_XO}
\end{figure}

To test the brightness of the host galaxy compared to the AGN, two criteria can be used. \\
Firstly, the X-ray to optical flux ratio ($\log{(f_X/f_{opt})}$ referred as the X/O ratio), which is usually used to discriminate between the different classes of X-ray sources \citep{Maccacaro88, DellaCeca04}. It can also be used to compare the AGN strength with that of its host galaxy.\\
The optical $r_{mag}$ is used to compute the optical flux:
$$f_{opt}=f_r=\Delta\lambda f_{\lambda,0} \cdot 10^{-0.4\cdot r_{mag}}$$
with $\Delta\lambda$ and  $f_{\lambda,0}$ from \citet{Fukugita95}) and $f_X$ corresponds to the hard X-ray flux.

As we can see in Figure \ref{fig:unobsSample_XO}, the \emph{unobscured Sy2} peak at lower X/O ratio than the \enquote{normal} obscured Sy2 (${f_X/f_{opt}}_{UnobsSy2} \sim0.041$ (i.e. $X/O_{UnobsSy2} \sim -1.4$), vs. ${f_X/f_{opt}}_{NormalSy2}\sim 0.36$, i.e. (i.e. $X/O_{NormalSy2} \sim -0.4$)). Also, a large fraction (68\%) of the \emph{unobscured Sy2} are found below the AGN locus ($-1<X/O<1$) while only 34\% of the normal Sy2 are outside this locus. The region where most of \emph{unobscured Sy2} are found (i.e. $X/O<-1$) is usually populated by normal galaxies or low-luminosity AGN \citep{Fiore03}. This result confirms the decreasing contrast between the AGN light and the host starlight for the \emph{unobscured Sy2}, suggesting the dilution hypothesis is possible if the BEL are weak.
\\
\begin{figure}
	\centering
	\includegraphics[width=\hsize]{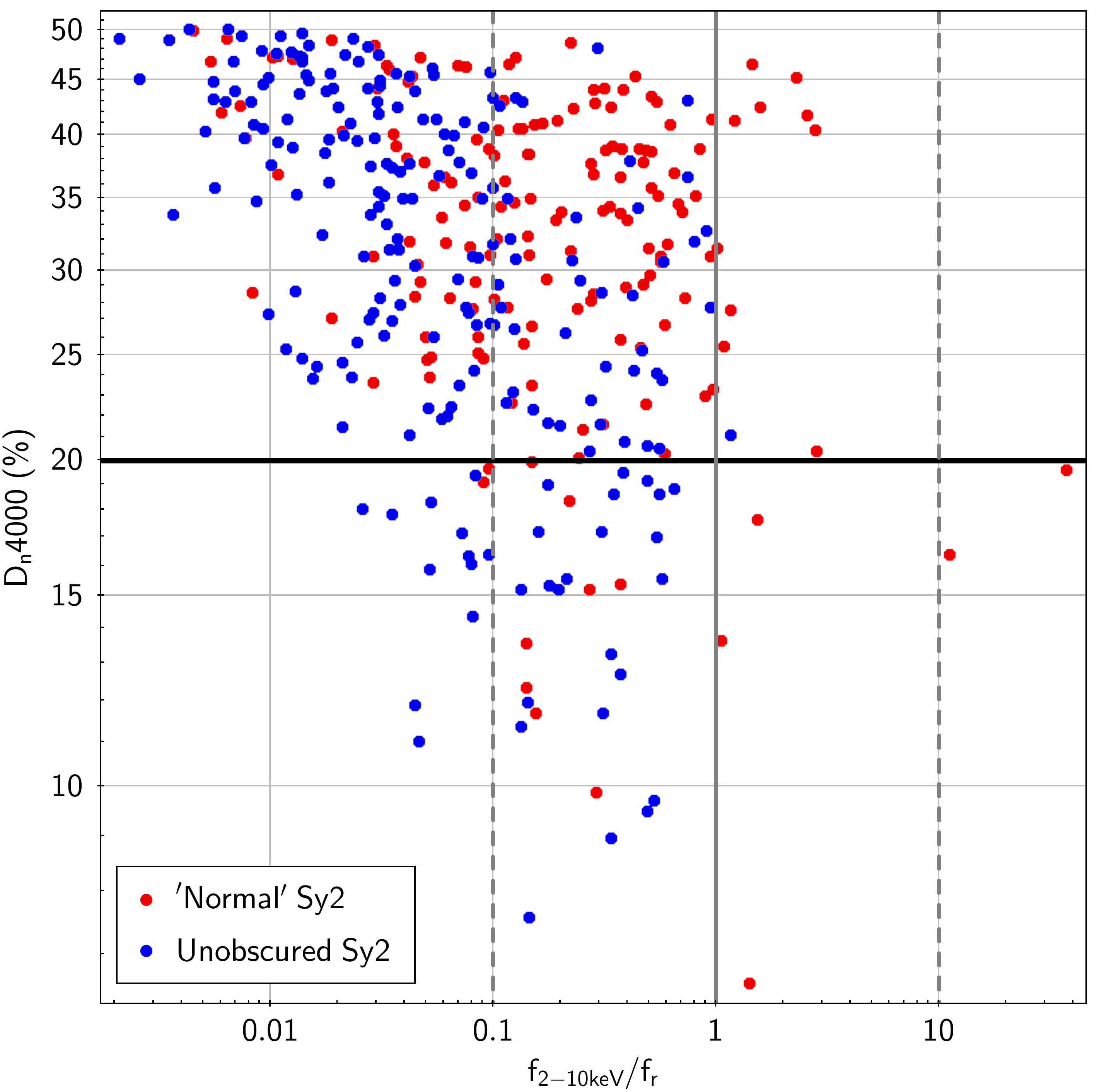}
	\caption[4000$\AAb$ break vs. X-ray to optical flux ratio for the normal Sy2and the \emph{unobscured Sy2}]{4000$\AAb$ break vs. X-ray to optical flux ratio for the normal Sy2 (red symbols) and the \emph{unobscured Sy2} (blue symbols). Dilution may be able to explain the absence of observed BEL, if they are weak, for galaxy dominated spectra (i.e. $D_n>20\%$; black line) with smaller contrast than usual between the AGN and the galaxy light (i.e. $f_X/f_{opt}<0.1$).}
	\label{fig:unobsSample_DnXO}
\end{figure}

Secondly, the 4000$\AAb$ break ($D_n=(F^+-F^-)/F^+$, $F^+$ and $F^-$ being respectively the mean flux density in the $4000-4100\AAb$ and $3850-3950\AAb$ rest-frame regions) allows the estimation of  the contribution to the optical continuum from the host galaxy. As found by \citet{Kriek06, Caccianiga07}, the galaxy will dominate the optical spectra for $D_n>20\%$ even if there can still be an AGN contribution for $D_n$ up to 60\%.

Both the normal Sy2 (91\%) and \emph{unobscured Sy2} (83\%) are mainly galaxy dominated as expected by \citet{Merloni14} and shown in Figure \ref{fig:unobsSample_DnXO}. Based on the $D_n$ and the X/O parameters, we can select the sources for which weak BEL may be diluted by the host galaxy, corresponding to $D_n>20\%$ in addition to $X/O<0.1$. Thus the BEL, if weak or very broad, may be overwhelmed by the host galaxy starlight for 62\% of the \emph{unobscured Sy2} sample.

	\subsection{True Sy2: sources which lack the BLR}
	\label{sec:no_BLR}
	
	\subsubsection{The disappearance of the BLR}
The last possible explanation to explain the absence of observed BEL (and the more interesting)  is that the BLR is in fact absent in these  AGN; these sources are the so-called \emph{True Sy2}. Theoretical works have predicted the disappearance of the BLR under a certain luminosity and/or accretion rates: in the context of a disk-wind models, \citet{Nicastro00} suggests that a vertical outflow is at the origin of the BLR clouds. Below certain accretion rates the accretion disk changes from radiation pressure dominated to gas-pressure dominated at a radius smaller than the innermost stable orbit, preventing the disk-wind from forming and thus the BLR to disappear.\\
Similarly, \citet{Elitzur06} and \citet{ElitzurHo09}, suggest that both the BLR and the torus are two parts of the same disk-driven wind: the inner hot ionised clouds are in fact the BLR while the outer clumpy dusty clouds form the torus. As the accretion rates drops below a critical value, the outflow is no longer supported, leading to the disappearance  of the torus and at some lower accretion rates to the vanishing of the BLR.\\
Alternatively, \citet{Laor03} suggested that there is a limit on the maximum velocity of the BEL due to the correlation between the size of the BLR and the luminosity. As such, for low-luminosity objects, the BLR shrinks and then disappears explaining why BEL are not observed in such objects.

In order to check if the \emph{unobscured Sy2} have properties consistent with such models we use the relation from \citet{Elitzur16} which is consistent with all disk outflow models and which predicts the disappearance of the BLR depending on the specific properties of the AGN, i.e. the BLR structure, the radiative efficiency and the fraction of the mass carried away by the outflow (parameterised by $\Lambda$, see eq. \ref{eq:RCh3_LbolnoBLR}). As these parameters vary from source to source, $\Lambda$ is also expected to vary among AGN.

\begin{equation}
	L_{bol, crit}(\Lambda) = \Lambda \left(\frac{M_{BH}}{10^7}\right)^{2/3} 
	\label{eq:RCh3_LbolnoBLR}
\end{equation}
with $\Lambda$ ranging from $\sim 4.7\times 10^{39}\;\ergsec$ to $\sim 3.5\times 10^{44}\;\ergsec$. The lower end of the $\Lambda$ parameter has been previously estimated directly from observations by \citet{ElitzurHo09} as they do not observe broad emission-lines in the AGN Palomar sample for bolometric luminosities smaller than $4.7\times10^{39}(M_{BH}/10^7)^{2/3}\;\ergsec$. On the contrary, the maximum value of $\Lambda$ has been inferred theoretically by \citet{Elitzur16} putting constraints on the mass conservation and luminosity of the AGN. The BLR is expected to vanish for $L_{bol}<L_{bol, crit}$, so for $L_{bol}< L_{bol, crit}(\Lambda_{min})$ it is certain that there is no BLR and the source can be securely classified as True Sy2. However, potential True Sy2 can also be found for $L_{bol, crit}(\Lambda_{min})<L_{bol}< L_{bol, crit}(\Lambda_{max})$ as this is a region where the BLR can be missing depending on the intrinsic properties of the AGN. Thus, AGN with the same BH mass could have different lower limits for the disappearance of the BLR.

\begin{figure}
	\centering
	\includegraphics[width=\hsize]{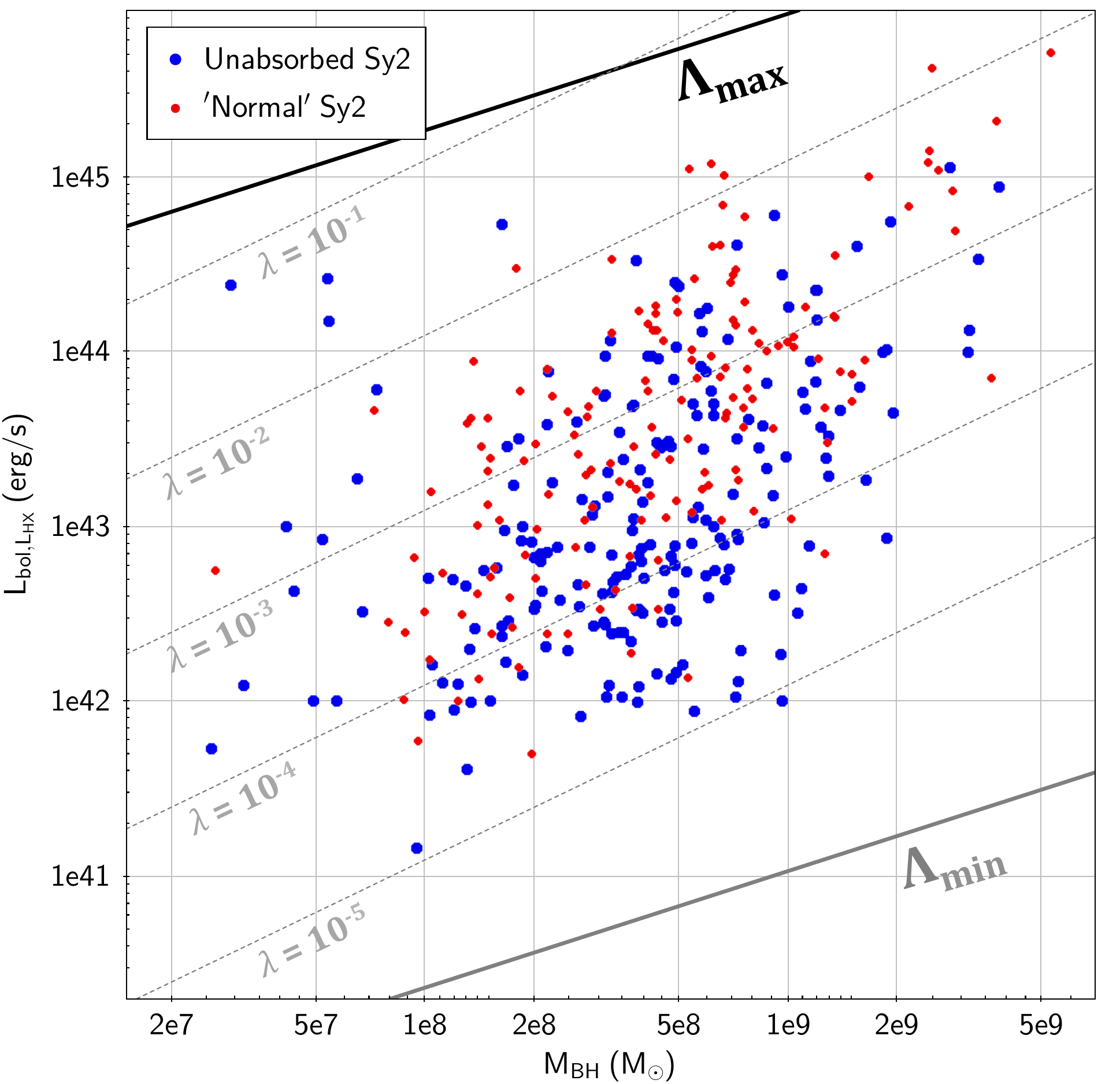}
	\caption[Bolometric vs. BH mass for the normal Sy2 and the \emph{unobscured Sy2}]{Bolometric vs. BH mass for the normal Sy2 (red symbols) in the \emph{unobscured Sy2} (blue symbols). All the  \emph{unobscured Sy2} are found in the same region as the potential True Sy2 (i.e. in the region between $\Lambda_{max}$, black line and $\Lambda_{min}$, grey line). $\Lambda_{max}$ and $\Lambda_{min}$ represent the limit for the existence of the BLR: below the minimal value we can be sure it is a True Sy2, AGN with an intrinsic BLR (even if not directly observed) are found above the maximal value, while between the two they can be either  True Sy2 or a BEL emitters.}
	\label{fig:unobsSample_LMBH}
\end{figure}

The bolometric luminosity and $M_{BH}$ are inferred using the equations \ref{eq:Lbol} and \ref{eq:MBH_LKtot}, described in section \ref{sec:var}.\\
As it can be seen in Figure \ref{fig:unobsSample_LMBH}, firstly, the \emph{unobscured Sy2} peak at slightly smaller Eddington ratio than the normal Sy2. They are all found in the potential True Sy2 region even when considering the large errors on the BH mass. However they cannot be securely confirmed as True Sy2 based only on  the bolometric luminosity and the BH mass alone as is discussed below.\\

	\subsubsection{The expected broad lines and the quality of the optical spectra}
We saw in the previous section (\ref{sec:unobsDil}), that 62\% of the sample may have their BEL diluted by the host galaxy. However this can only be the case if the lines are weak or so broad that they are hidden in the noise of the spectra. So in order to find a secure sample of True Sy2, the strength and width of the BEL needs to be assessed and compared with the optical spectra.

\begin{figure}
	\centering
	\includegraphics[width=\hsize]{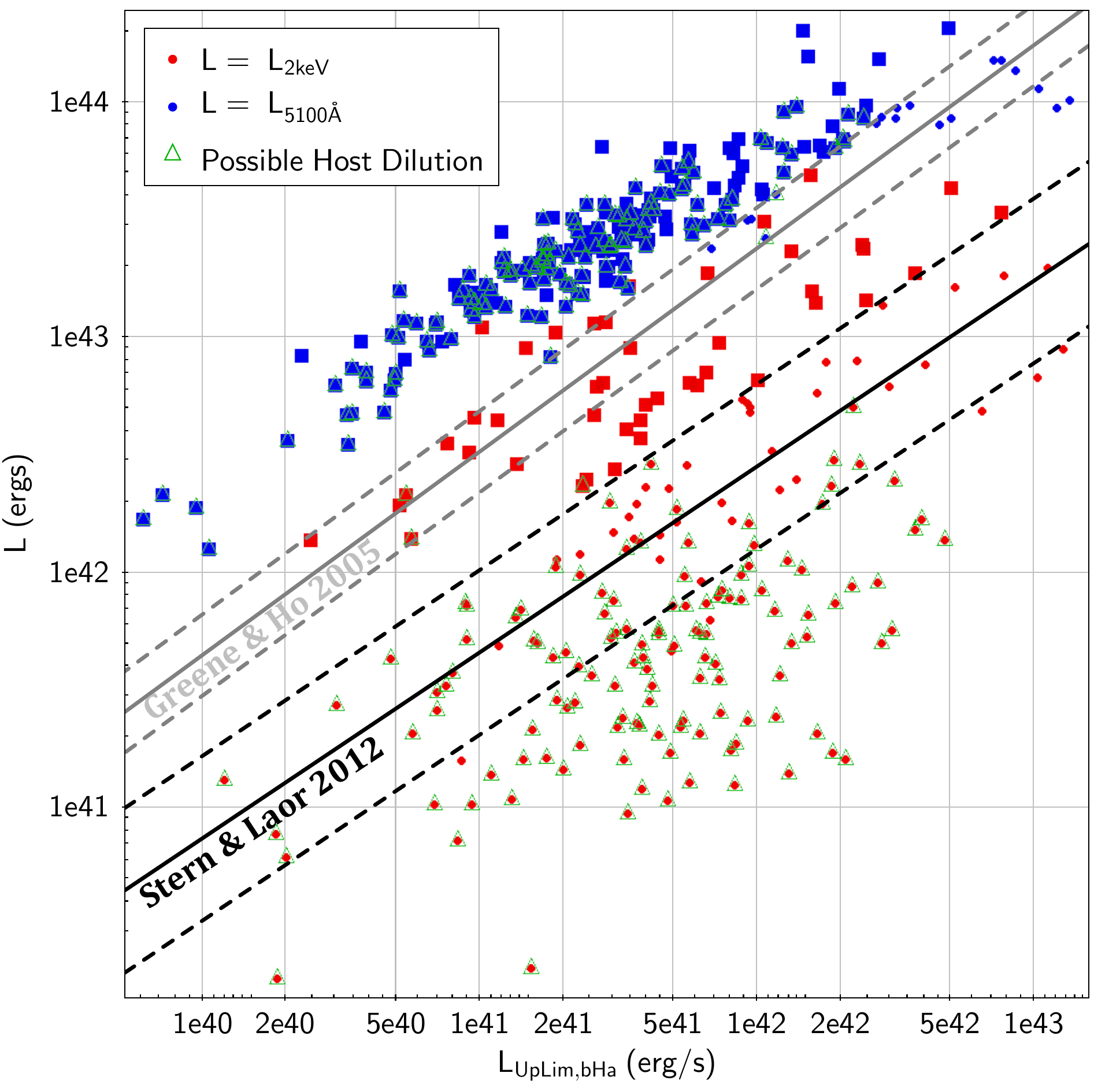}
	\caption{AGN luminosity ($L_{5100\AAb}$ for blue symbols and $L_{2\kev}$ for red symbols) vs. the luminosity of the broad H$\alpha$. The grey line represents the relation of \citet{Greene05} with the dotted lines corresponding to the 0.2 dex scatter, while the black line is for the \citet{SternLaor12} relation with a scatter of 0.35 dex. The sources for which the BEL would have been securely detected if present are labelled by squares. While this is the case for 91\% (blue squares) of the sample based on $L_{5100\AAb}$, from  $L_{2\kev}$ it is only 21\% (red squares). Sources that may have their BEL diluted by the host galaxy are labelled by open green triangles.}
	\label{fig:unobsSample_LLbHa}
\end{figure}

A BEL will be securely detected in the optical spectra if it lies above the noise of the spectra, and thus we define the detection criterion (assuming that the emission-lines are Gaussian) as:
$$f_{exp, BL} > f_{UpLim, BL} = \frac{1}{2} \sqrt{\left(\frac{\pi}{\ln{2}}\right)} \cdot 3\sigma \cdot FWHM_{exp, BL}$$
with $f_{BL,exp}$ (in units of $\ergsec$) being the expected flux of the broad component, $FWHM_{exp, BL}$ the expected width of the broad component (in units of $\AAb$) and $\sigma$ the noise in the spectra around the broad line (in units of $\ergcmsA$).\\
In previous work \citep{PW14, Pons16b}, we have used the luminosity at $L_{5100\AAb}$ to estimate the expected luminosity \citep[relation from][see eq. \ref{eq:BL_LbHa}]{Greene05} and width of the broad component of the Balmer lines \citep[relation from][see eq.\ref{eq:BL_FWHM}]{Xiao11}:
\begin{equation}
	L_{Broad\,H\alpha} =(5.25\pm 0.02)\cdot 10^{42}\left(\frac{\mathbf{\lambda}L_{5100}}{10^{44}\,\ergsec}\right)^{(1.157\pm 0.005)} 	\label{eq:BL_LbHa}
\end{equation}
\begin{equation}
	M_{BH}\,(\mathrm{M_{\odot}})=3.47\cdot \left[\frac{\mathbf{\lambda}L_{5100\AA}}{10^{44}\, \mathrm{erg.s^{-1}}}\right]^{0.519}\cdot \left[\frac{FWHM_{H\alpha}}{\mathrm{km.s^{-1}}}\right]^{2.06}
	\label{eq:BL_FWHM}
\end{equation}
This is the same method used by \citet{Ho12, Miniutti13} to confirm the absence of BEL in their sources.

However, more recent work by \citet{SternLaor12} has found a correlation between the rest-frame X-ray luminosity at $2\;\kev$ ($L_{2\kev}$) and the width and luminosity of the broad H$\alpha$ line based on a large sample of type-1 AGN from SDSS:

\begin{equation}
	\log{\left(\frac{L_{2\kev}}{10^{42}\;\ergsec}\right)}=0.79\; \log{\left(\frac{L_{b, H\alpha}}{10^{42}\;\ergsec}\right)}
	\label{eq:RCh3_BL_L2LbHa}
\end{equation}
with an intrinsic dispersion of about $0.35\;\mathrm{dex}$ and
\begin{multline}
	\log{\left(\frac{M_{BH}}{\Msun}\right)}=7.4+2.06\;\log{\left(\frac{FWHM_{b, H\alpha}}{1000\;\kms}\right)}\\
	+0.545\;\log{\left(\frac{L_{b, H\alpha}}{10^{44}\;\ergsec}\right)}
	\label{eq:RCh3_BL_MBHLbHa}
\end{multline}

Figure \ref{fig:unobsSample_LLbHa} shows the upper limit on the broad H$\alpha$ luminosity from the \citet[i.e. from the $L_{5100\AAb}$]{Greene05} and \citet[i.e. from the $L_{2\kev}$]{SternLaor12} relations. We can consider that the BEL would have have securely detected if the upper limit for the broad H$\alpha$ luminosity is smaller than the expected value (even when the scatter is considered) based on the AGN luminosity (at $L_{5100\AAb}$ or $L_{2\kev}$), and thus in this case the dilution will be unable to explain the absence of BEL. It is however important to remember that even if the $L_{bH\alpha}$ upper limit is larger than the expected value, it does not mean that the broad H$\alpha$ must be there, just that the quality of the data is not conclusive to demonstrate its absence.\\
We clearly see that, while the majority of the \emph{unobscured Sy2} (91\%) would have their BEL securely detected if present assuming the $L_{5100\AAb}-L_{Broad,H\alpha}$ relation, only 21\% (i.e. 44 sources) of the \emph{unobscured Sy2} would have their BEL securely detected from the $L_{2\kev}-L_{Broad,H\alpha}$ relation. 

To understand the difference, we can look at the difference between the derived parameters from the two relations. Firstly, the expected FWHM of the broad H$\alpha$ from $L_{5100\AAb}$ is globally smaller than from $L_{2\kev}$, but the two measurements agree (within $\pm 10\%$) when the large errors are considered. On the contrary, the expected broad H$\alpha$ luminosity from $L_{5100\AAb}$ is generally larger than the predicted value from $L_{2\kev}$ and differs by more than 10\% for half of the sample even when considering the scatter in the relations.\\
Moreover, from the Figure 10 of \citet{SternLaor12} ($L_{5100\AAb}$ vs $L_{b,H\alpha}$), the relation from \citet{Greene05} does not fit the type-1 SDSS sample of \citet{SternLaor12} for low 5100$\AAb$ luminosity ($L_{5100\AAb} \lesssim 4\times 10^{43}\; \ergsec$) and overestimates the broad H$\alpha$ luminosity compared with what is observed. As most (about 70\%) of our \emph{unobscured Sy2} have $L_{5100\AAb} < 4\times 10^{43}\; \ergsec$, it may explain why $L_{b,H\alpha}(L_{5100\AAb})> L_{b,H\alpha}(L_{2\kev})$.

In order to have a secure  selection we consider the results from the $L_{2\kev}$ relation of \citet{SternLaor12}. First of all, 44 sources among the 207 \emph{unobscured Sy2} (i.e. 21\%) would have their broad H$\alpha$ line detected if present ($L_{UpLim, bH\alpha} < L_{exp, bH\alpha}$ by at least 0.35 dex). Host dilution is unable to explain the absence of BEL in this case and the extinction by large scale dust is very unlikely as the predicted broad H$\alpha$ luminosity for these sources is higher than the luminosity of the observed narrow H$\alpha$ component. Due to the delay between the \xmm and SDSS observations, and because of the absence of several X-ray observations or optical data sets, variability cannot be ruled out for 11 of the 44 sources and thus \enquote{simultaneous} X-ray and optical observations will be necessary to confirm the classification as True Sy2.\\
For the 163 remaining sources with $L_{UpLim, bH\alpha} > L_{exp, bH\alpha}$ (or $L_{UpLim, bH\alpha} < L_{exp, bH\alpha}$ by less than 0.35 dex), host dilution is the best explanation to explain the absence of BEL for 77\% of them (see Figure \ref{fig:unobsSample_LLbHa}). As above, dust extinction is unlikely as $L_{exp,bH\alpha} \gtrsim L_{obs, nH\alpha}$. Only one source clearly has a variable X-ray HR, that could explain the absence of X-ray obscuration and for 84 others variability cannot be excluded. It is also possible that the quality of the optical data is not sufficient to detect the BEL or these are indeed True Sy2.


\section{Discussion and Conclusion}
From the 3XMM-SDSS cross-matched sample, we have selected NL AGN based on both X-ray and optical criteria; we found that AGN represent about 60\% of the initial sample. Then, we have looked at the level of X-ray obscuration for these sources. As only half the sources of the sample have available X-ray spectra with a fit already performed by \citet{Corral15}, we have estimated the reliability of the X-ray HR to distinguish between obscured and unobscured sources as it is an approximate indicator of the shape of the X-ray spectrum and it is often used as an indicator of the level of absorption when X-ray spectra have too low a number of counts or are not available. We found that there is in fact a very small fraction (only 2\%) of the sources classified as unobscured by the HR which have $N_H>4\times 10^{21}\;\cmsq$ and only 4\% the HR obscured sources with an absence of absorption in the X-ray spectra. Thus, the HR can be considered as a highly reliable parameter to select unobscured AGN.

From the parent \emph{NL AGN} sample, 279 sources (63\%) were classified as unobscured but this includes objects misclassified as NL (i.e. NLS1) and those misclassified as unobscured (i.e. CT AGN). By looking at the Balmer vs. forbidden line width and at the optical spectra we were able to find 44 NLS1. In addition, two indirect measurements of the AGN luminosity have been compared ($L_{HX}/L_{[OIII]}$ and $L_{HX}/L_{12\mu m}$) to identify heavily obscured sources, and 28 CT AGN were found. \\
This leads to an uncontaminated \emph{X-ray unobscured Sy2} sample of 207 sources corresponding to 47\% of the parent \emph{NL AGN} sample. This is a quite high fraction compared to the 30\% found by \citet{Merloni14} for the \emph{XMM}-COSMOS field but still consistent with the fraction of 40\% claimed by \citet{Marinucci12}. So our results may be in the range found by other authors, and this seems to suggest that unobscured Sy2 may be a significant part of the AGN population.

Based on disc outflow models, the BLR disappears below a critical bolometric luminosity which depends on the black hole mass. Indeed all the unobscured Sy2 of our sample can be considered as potential True Sy2 from their bolometric luminosity and BH mass (i.e. the BLR may be absent in these sources).

While \citet{Merloni14}, suggested that the dilution by the host galaxy will be the most plausible explanation, an important parameter to take into account is the the ratio between the expected broad component luminosity and the upper limit luminosity based on the noise of the optical spectra. Indeed dilution may be able to explain the absence of observed BEL but only if the upper limit luminosity is larger than the expected luminosity for the broad component; only if this is the case may weak BEL or very broad emission-lines be hidden in the continuum and/or noise of the optical spectra.\\
It is important to note that the secure detection of the BEL depends on which AGN luminosity used to derive the expected FWHM and luminosity of the broad component. We found that the expected luminosity of the broad H$\alpha$ line predicted by $L_{5100\AAb}$ \citep[relation from][]{Greene05} is up to 10 times larger than the luminosity predicted by $L_{2\kev}$ \citep[relation from][]{SternLaor12}. This leads to a larger number of sources with $L_{UpLim, bH\alpha}< L_{exp,bH\alpha}$ from $L_{5100\AAb}$ (91\% of the \emph{unobscured Sy2} sample) that from $L_{2\kev}$ (21\%).\\
However, based on the data sets of \citet{SternLaor12}, the \citet{Greene05} relation overestimated the luminosity of the broad H$\alpha$ component for $L_{5100\AAb} < \sim 4\times 10^{43} \;\ergsec$ and the luminosity at 5100$\AAb$ for our sample is mainly below this threshold.

To have a secure estimate of the number of True Sy2 in our \emph{unobscured Sy2}, we used the results from the luminosity at 2 $\kev$. For the 44 sources with $L_{UpLim, bH\alpha}< L_{exp,bH\alpha}$, the only other possible explanation other than the True Sy2 classification for our sample is variability. Indeed, this cannot be ruled out for 11 sources and these will require simultaneous X-ray and optical observations to confirm their True Sy2 classification. Even the lower limit of 33 sources is about three times the number of confirmed True Sy2.\\
For the other sources with $L_{UpLim, bH\alpha}>L_{exp,bH\alpha}$, dilution by the host galaxy is the most plausible explanation for 77\% of them, but variability, insufficient S/N optical spectra and True Sy2 may be able to explain the absence of BEL for the remaining 33\%.\\
Thus the lower limit on the fraction of True Sy2 is 16\% of the \emph{unobscured Sy2} sample, corresponding to 7\% of the parent \emph{NL AGN} sample and suggesting that the True Sy2 may be a non-negligible fraction of the AGN population which could have important implication for the UM. Indeed, the viewing angle seems to be not the only parameter to distinguish between type-1 and type-2 AGN and the accretion rates of the AGN may play an important role in the classification.	\\

Also, the distinction between unobscured and obscured AGN plays an important role in the understanding of the X-ray Background (XRB). Indeed the population synthesis model from \citet{Gilli01} predicts a ratio between absorbed and unabsorbed AGN ranging from 4 (at $z=0$) to 10 (at $z=1.3$) to fit the hard X-ray peak in the XRB spectrum. However, this model is based on the assumption that optical and X-ray classification agrees, with the fraction of obscured AGN being inferred from the optically type-2 (i.e. narrow-line) population. As we saw in this work, about half of the narrow-line AGN are indeed X-ray unobscured; and thus the estimated fraction of obscured AGN (from optical based classification) in XRB synthesis models will be overestimated. Then in order to fit the hard spectrum of the XRB, some obscured AGN are still missing: they may be Compton-thick AGN, high-z type-2 quasars or either optically \emph{elusive} AGN which would have escaped optical spectroscopic observations/identification due to their high column densities, weak optical counterparts or the lack of AGN optical signature.

\begin{acknowledgements}
This work uses data from observations obtained with XMM-Newton, an ESA science mission with instruments and contributions directly funded by ESA Member States and NASA. Data from the SDSS is also utilised. Funding for SDSS-III has been provided by the Alfred P. Sloan Foundation, the Participating Institutions, the National Science Foundation, and the U.S. Department of Energy Office of Science. The SDSS-III web site is http://www.sdss3.org/. 
\end{acknowledgements}


\bibliographystyle{aa}
\bibliography{bib_pIV.bib}

\begin{thebibliography}{77}
\expandafter\ifx\csname natexlab\endcsname\relax\def\natexlab#1{#1}\fi

\bibitem[{{Alexander} {et~al.}(2008){Alexander}, {Chary}, {Pope}, {Bauer},
  {Brandt}, {Daddi}, {Dickinson}, {Elbaz}, \& {Reddy}}]{Alexander08}
{Alexander}, D.~M., {Chary}, R.-R., {Pope}, A., {et~al.} 2008, \apj, 687, 835

\bibitem[{{Alonso-Herrero} {et~al.}(1996){Alonso-Herrero}, {Ward}, \&
  {Kotilainen}}]{AWK96}
{Alonso-Herrero}, A., {Ward}, M.~J., \& {Kotilainen}, J.~K. 1996, \mnras, 278,
  902

\bibitem[{{Antonucci}(1993)}]{Antonucci93}
{Antonucci}, R. 1993, \araa, 31, 473

\bibitem[{{Antonucci} \& {Olszewski}(1985)}]{Antonucci85}
{Antonucci}, R.~R.~J. \& {Olszewski}, E.~W. 1985, \aj, 90, 2203

\bibitem[{{Baldwin} {et~al.}(1981){Baldwin}, {Phillips}, \&
  {Terlevich}}]{BPT81}
{Baldwin}, J.~A., {Phillips}, M.~M., \& {Terlevich}, R. 1981, \pasp, 93, 5

\bibitem[{{Bassani} {et~al.}(1999){Bassani}, {Dadina}, {Maiolino}, {Salvati},
  {Risaliti}, {della Ceca}, {Matt}, \& {Zamorani}}]{Bassani99}
{Bassani}, L., {Dadina}, M., {Maiolino}, R., {et~al.} 1999, \apjs, 121, 473

\bibitem[{{Bianchi} {et~al.}(2012{\natexlab{a}}){Bianchi}, {Maiolino}, \&
  {Risaliti}}]{Bianchi12b}
{Bianchi}, S., {Maiolino}, R., \& {Risaliti}, G. 2012{\natexlab{a}}, Advances
  in Astronomy, 2012, 782030

\bibitem[{{Bianchi} {et~al.}(2012{\natexlab{b}}){Bianchi}, {Panessa},
  {Barcons}, {Carrera}, {La Franca}, {Matt}, {Onori}, {Wolter}, {Corral},
  {Monaco}, {Ruiz}, \& {Brightman}}]{Bianchi12}
{Bianchi}, S., {Panessa}, F., {Barcons}, X., {et~al.} 2012{\natexlab{b}},
  \mnras, 426, 3225

\bibitem[{{Bianchi} {et~al.}(2009){Bianchi}, {Piconcelli}, {Chiaberge},
  {Bail{\'o}n}, {Matt}, \& {Fiore}}]{BianchiPiconcelli09}
{Bianchi}, S., {Piconcelli}, E., {Chiaberge}, M., {et~al.} 2009, \apj, 695, 781

\bibitem[{{Bongiorno} {et~al.}(2012){Bongiorno}, {Merloni}, {Brusa},
  {Magnelli}, {Salvato}, {Mignoli}, {Zamorani}, {Fiore}, {Rosario}, {Mainieri},
  {Hao}, {Comastri}, {Vignali}, {Balestra}, {Bardelli}, {Berta}, {Civano},
  {Kampczyk}, {Le Floc'h}, {Lusso}, {Lutz}, {Pozzetti}, {Pozzi}, {Riguccini},
  {Shankar}, \& {Silverman}}]{Bongiorno12}
{Bongiorno}, A., {Merloni}, A., {Brusa}, M., {et~al.} 2012, \mnras, 427, 3103

\bibitem[{{Brightman} \& {Nandra}(2008)}]{BrightmanNandra08}
{Brightman}, M. \& {Nandra}, K. 2008, \mnras, 390, 1241

\bibitem[{{Brightman} \& {Nandra}(2011)}]{BrightmanNandra11}
{Brightman}, M. \& {Nandra}, K. 2011, \mnras, 414, 3084

\bibitem[{{Brinchmann} {et~al.}(2004){Brinchmann}, {Charlot}, {White},
  {Tremonti}, {Kauffmann}, {Heckman}, \& {Brinkmann}}]{Brinchmann04}
{Brinchmann}, J., {Charlot}, S., {White}, S.~D.~M., {et~al.} 2004, \mnras, 351,
  1151

\bibitem[{{Brusa} {et~al.}(2003){Brusa}, {Comastri}, {Mignoli}, {Fiore},
  {Ciliegi}, {Vignali}, {Severgnini}, {Cocchia}, {La Franca}, {Matt}, {Perola},
  {Maiolino}, {Baldi}, \& {Molendi}}]{Brusa03}
{Brusa}, M., {Comastri}, A., {Mignoli}, M., {et~al.} 2003, \aap, 409, 65

\bibitem[{{Caccianiga} {et~al.}(2004){Caccianiga}, {Severgnini}, {Braito},
  {Della Ceca}, {Maccacaro}, {Wolter}, {Barcons}, {Carrera}, {Lehmann}, {Page},
  {Saxton}, \& {Webb}}]{Caccianiga04}
{Caccianiga}, A., {Severgnini}, P., {Braito}, V., {et~al.} 2004, \aap, 416, 901

\bibitem[{{Caccianiga} {et~al.}(2007){Caccianiga}, {Severgnini}, {Della Ceca},
  {Maccacaro}, {Carrera}, \& {Page}}]{Caccianiga07}
{Caccianiga}, A., {Severgnini}, P., {Della Ceca}, R., {et~al.} 2007, \aap, 470,
  557

\bibitem[{{Caccianiga} {et~al.}(2008){Caccianiga}, {Severgnini}, {Della Ceca},
  {Maccacaro}, {Cocchia}, {Barcons}, {Carrera}, {Matute}, {McMahon}, {Page},
  {Pietsch}, {Sbarufatti}, {Schwope}, {Tedds}, \& {Watson}}]{Caccianiga08}
{Caccianiga}, A., {Severgnini}, P., {Della Ceca}, R., {et~al.} 2008, \aap, 477,
  735

\bibitem[{{Castell{\'o}-Mor} {et~al.}(2012){Castell{\'o}-Mor}, {Barcons},
  {Ballo}, {Carrera}, {Ward}, \& {Jin}}]{Castello12}
{Castell{\'o}-Mor}, N., {Barcons}, X., {Ballo}, L., {et~al.} 2012, \aap, 544,
  A48

\bibitem[{{Chary} \& {Elbaz}(2001)}]{Chary01}
{Chary}, R. \& {Elbaz}, D. 2001, \apj, 556, 562

\bibitem[{{Corral} {et~al.}(2015){Corral}, {Georgantopoulos}, {Watson},
  {Rosen}, {Page}, \& {Webb}}]{Corral15}
{Corral}, A., {Georgantopoulos}, I., {Watson}, M.~G., {et~al.} 2015, \aap, 576,
  A61

\bibitem[{{Della Ceca} {et~al.}(2004){Della Ceca}, {Maccacaro}, {Caccianiga},
  {Severgnini}, {Braito}, {Barcons}, {Carrera}, {Watson}, {Tedds}, {Brunner},
  {Lehmann}, {Page}, {Lamer}, \& {Schwope}}]{DellaCeca04}
{Della Ceca}, R., {Maccacaro}, T., {Caccianiga}, A., {et~al.} 2004, \aap, 428,
  383

\bibitem[{{Elitzur} \& {Ho}(2009)}]{ElitzurHo09}
{Elitzur}, M. \& {Ho}, L.~C. 2009, \apjl, 701, L91

\bibitem[{{Elitzur} \& {Netzer}(2016)}]{Elitzur16}
{Elitzur}, M. \& {Netzer}, H. 2016, \mnras, 459, 585

\bibitem[{{Elitzur} \& {Shlosman}(2006)}]{Elitzur06}
{Elitzur}, M. \& {Shlosman}, I. 2006, \apjl, 648, L101

\bibitem[{{Elvis} {et~al.}(2004){Elvis}, {Risaliti}, {Nicastro}, {Miller},
  {Fiore}, \& {Puccetti}}]{Elvis04}
{Elvis}, M., {Risaliti}, G., {Nicastro}, F., {et~al.} 2004, \apjl, 615, L25

\bibitem[{{Fiore} {et~al.}(2003){Fiore}, {Brusa}, {Cocchia}, {Baldi},
  {Carangelo}, {Ciliegi}, {Comastri}, {La Franca}, {Maiolino}, {Matt},
  {Molendi}, {Mignoli}, {Perola}, {Severgnini}, \& {Vignali}}]{Fiore03}
{Fiore}, F., {Brusa}, M., {Cocchia}, F., {et~al.} 2003, \aap, 409, 79

\bibitem[{{Fukugita} {et~al.}(1995){Fukugita}, {Shimasaku}, \&
  {Ichikawa}}]{Fukugita95}
{Fukugita}, M., {Shimasaku}, K., \& {Ichikawa}, T. 1995, \pasp, 107, 945

\bibitem[{{Gandhi} {et~al.}(2009){Gandhi}, {Horst}, {Smette}, {H{\"o}nig},
  {Comastri}, {Gilli}, {Vignali}, \& {Duschl}}]{Gandhi09}
{Gandhi}, P., {Horst}, H., {Smette}, A., {et~al.} 2009, \aap, 502, 457

\bibitem[{{Garcet} {et~al.}(2007){Garcet}, {Gandhi}, {Gosset}, {Sprimont},
  {Surdej}, {Borkowski}, {Tajer}, {Pacaud}, {Pierre}, {Chiappetti}, {Maccagni},
  {Page}, {Carrera}, {Tedds}, {Mateos}, {Krumpe}, {Contini}, {Corral},
  {Ebrero}, {Gavignaud}, {Schwope}, {Le F{\`e}vre}, {Polletta}, {Rosen},
  {Lonsdale}, {Watson}, {Borczyk}, \& {Vaisanen}}]{Garcet07}
{Garcet}, O., {Gandhi}, P., {Gosset}, E., {et~al.} 2007, \aap, 474, 473

\bibitem[{{Georgantopoulos} {et~al.}(2011){Georgantopoulos}, {Rovilos},
  {Akylas}, {Comastri}, {Ranalli}, {Vignali}, {Balestra}, {Gilli}, \&
  {Cappelluti}}]{Georgantopoulos11}
{Georgantopoulos}, I., {Rovilos}, E., {Akylas}, A., {et~al.} 2011, \aap, 534,
  A23

\bibitem[{{Gilli} {et~al.}(2001){Gilli}, {Salvati}, \& {Hasinger}}]{Gilli01}
{Gilli}, R., {Salvati}, M., \& {Hasinger}, G. 2001, \aap, 366, 407

\bibitem[{{Goulding} {et~al.}(2011){Goulding}, {Alexander}, {Mullaney},
  {Gelbord}, {Hickox}, {Ward}, \& {Watson}}]{Goulding11}
{Goulding}, A.~D., {Alexander}, D.~M., {Mullaney}, J.~R., {et~al.} 2011,
  \mnras, 411, 1231

\bibitem[{{Greene} \& {Ho}(2005)}]{Greene05}
{Greene}, J.~E. \& {Ho}, L.~C. 2005, \apj, 630, 122

\bibitem[{{G{\"u}ver} \& {{\"O}zel}(2009)}]{Guver09}
{G{\"u}ver}, T. \& {{\"O}zel}, F. 2009, \mnras, 400, 2050

\bibitem[{{Hawkins}(2004)}]{Hawkins04}
{Hawkins}, M.~R.~S. 2004, \aap, 424, 519

\bibitem[{{Ho} {et~al.}(2012){Ho}, {Kim}, \& {Terashima}}]{Ho12}
{Ho}, L.~C., {Kim}, M., \& {Terashima}, Y. 2012, \apjl, 759, L16

\bibitem[{{Huang} {et~al.}(2011){Huang}, {Wang}, {Tan}, {Yang}, \&
  {Huang}}]{Huang11}
{Huang}, X.-X., {Wang}, J.-X., {Tan}, Y., {Yang}, H., \& {Huang}, Y.-F. 2011,
  \apjl, 734, L16

\bibitem[{{Kauffmann} {et~al.}(2003{\natexlab{a}}){Kauffmann}, {Heckman},
  {Tremonti}, {Brinchmann}, {Charlot}, {White}, {Ridgway}, {Brinkmann},
  {Fukugita}, {Hall}, {Ivezi{\'c}}, {Richards}, \& {Schneider}}]{Kauff03b}
{Kauffmann}, G., {Heckman}, T.~M., {Tremonti}, C., {et~al.} 2003{\natexlab{a}},
  \mnras, 346, 1055

\bibitem[{{Kauffmann} {et~al.}(2003{\natexlab{b}}){Kauffmann}, {Heckman},
  {White}, {Charlot}, {Tremonti}, {Brinchmann}, {Bruzual}, {Peng}, {Seibert},
  {Bernardi}, {Blanton}, {Brinkmann}, {Castander}, {Cs{\'a}bai}, {Fukugita},
  {Ivezic}, {Munn}, {Nichol}, {Padmanabhan}, {Thakar}, {Weinberg}, \&
  {York}}]{Kauff03a}
{Kauffmann}, G., {Heckman}, T.~M., {White}, S.~D.~M., {et~al.}
  2003{\natexlab{b}}, \mnras, 341, 33

\bibitem[{{Kewley} {et~al.}(2001){Kewley}, {Dopita}, {Sutherland}, {Heisler},
  \& {Trevena}}]{Kewley01}
{Kewley}, L.~J., {Dopita}, M.~A., {Sutherland}, R.~S., {Heisler}, C.~A., \&
  {Trevena}, J. 2001, \apj, 556, 121

\bibitem[{{Kriek} {et~al.}(2006){Kriek}, {van Dokkum}, {Franx}, {F{\"o}rster
  Schreiber}, {Gawiser}, {Illingworth}, {Labb{\'e}}, {Marchesini}, {Quadri},
  {Rix}, {Rudnick}, {Toft}, {van der Werf}, \& {Wuyts}}]{Kriek06}
{Kriek}, M., {van Dokkum}, P.~G., {Franx}, M., {et~al.} 2006, \apj, 645, 44

\bibitem[{{Laor}(2003)}]{Laor03}
{Laor}, A. 2003, \apj, 590, 86

\bibitem[{{L{\"a}sker} {et~al.}(2014){L{\"a}sker}, {Ferrarese}, {van de Ven},
  \& {Shankar}}]{Lasker14}
{L{\"a}sker}, R., {Ferrarese}, L., {van de Ven}, G., \& {Shankar}, F. 2014,
  \apj, 780, 70

\bibitem[{{Maccacaro} {et~al.}(1988){Maccacaro}, {Gioia}, {Wolter}, {Zamorani},
  \& {Stocke}}]{Maccacaro88}
{Maccacaro}, T., {Gioia}, I.~M., {Wolter}, A., {Zamorani}, G., \& {Stocke},
  J.~T. 1988, \apj, 326, 680

\bibitem[{{Mainieri} {et~al.}(2002){Mainieri}, {Bergeron}, {Hasinger},
  {Lehmann}, {Rosati}, {Schmidt}, {Szokoly}, \& {Della Ceca}}]{Mainieri02}
{Mainieri}, V., {Bergeron}, J., {Hasinger}, G., {et~al.} 2002, \aap, 393, 425

\bibitem[{{Maiolino} {et~al.}(2001){Maiolino}, {Marconi}, {Salvati},
  {Risaliti}, {Severgnini}, {Oliva}, {La Franca}, \& {Vanzi}}]{Maiolino01}
{Maiolino}, R., {Marconi}, A., {Salvati}, M., {et~al.} 2001, \aap, 365, 28

\bibitem[{{Marconi} {et~al.}(2004){Marconi}, {Risaliti}, {Gilli}, {Hunt},
  {Maiolino}, \& {Salvati}}]{Marconi04}
{Marconi}, A., {Risaliti}, G., {Gilli}, R., {et~al.} 2004, \mnras, 351, 169

\bibitem[{{Marinucci} {et~al.}(2012){Marinucci}, {Bianchi}, {Nicastro}, {Matt},
  \& {Goulding}}]{Marinucci12}
{Marinucci}, A., {Bianchi}, S., {Nicastro}, F., {Matt}, G., \& {Goulding},
  A.~D. 2012, \apj, 748, 130

\bibitem[{{Merloni} {et~al.}(2014){Merloni}, {Bongiorno}, {Brusa}, {Iwasawa},
  {Mainieri}, {Magnelli}, {Salvato}, {Berta}, {Cappelluti}, {Comastri},
  {Fiore}, {Gilli}, {Koekemoer}, {Le Floc'h}, {Lusso}, {Lutz}, {Miyaji},
  {Pozzi}, {Riguccini}, {Rosario}, {Silverman}, {Symeonidis}, {Treister},
  {Vignali}, \& {Zamorani}}]{Merloni14}
{Merloni}, A., {Bongiorno}, A., {Brusa}, M., {et~al.} 2014, \mnras, 437, 3550

\bibitem[{{Miniutti} {et~al.}(2013){Miniutti}, {Saxton},
  {Rodr{\'{\i}}guez-Pascual}, {Read}, {Esquej}, {Colless}, {Dobbie}, \&
  {Spolaor}}]{Miniutti13}
{Miniutti}, G., {Saxton}, R.~D., {Rodr{\'{\i}}guez-Pascual}, P.~M., {et~al.}
  2013, \mnras, 433, 1764

\bibitem[{{Nicastro}(2000)}]{Nicastro00}
{Nicastro}, F. 2000, \apjl, 530, L65

\bibitem[{{Page} {et~al.}(2006){Page}, {Loaring}, {Dwelly}, {Mason}, {McHardy},
  {Gunn}, {Moss}, {Sasseen}, {Cordova}, {Kennea}, \& {Seymour}}]{Page06}
{Page}, M.~J., {Loaring}, N.~S., {Dwelly}, T., {et~al.} 2006, \mnras, 369, 156

\bibitem[{{Panessa} \& {Bassani}(2002)}]{PanessaBassani02}
{Panessa}, F. \& {Bassani}, L. 2002, \aap, 394, 435

\bibitem[{{Pineau} {et~al.}(2011){Pineau}, {Motch}, {Carrera}, {Della Ceca},
  {Derri{\`e}re}, {Michel}, {Schwope}, \& {Watson}}]{Pineau11}
{Pineau}, F.-X., {Motch}, C., {Carrera}, F., {et~al.} 2011, \aap, 527, A126

\bibitem[{{Pogge}(2000)}]{Pogge00}
{Pogge}, R.~W. 2000, \nar, 44, 381

\bibitem[{{Pons} {et~al.}(2016){Pons}, {Elvis}, {Civano}, \&
  {Watson}}]{Pons16b}
{Pons}, E., {Elvis}, M., {Civano}, F., \& {Watson}, M.~G. 2016, \apj, in press

\bibitem[{{Pons} \& {Watson}(2014)}]{PW14}
{Pons}, E. \& {Watson}, M.~G. 2014, \aap, 568, A108

\bibitem[{{Rigby} {et~al.}(2006){Rigby}, {Rieke}, {Donley}, {Alonso-Herrero},
  \& {P{\'e}rez-Gonz{\'a}lez}}]{Rigby06}
{Rigby}, J.~R., {Rieke}, G.~H., {Donley}, J.~L., {Alonso-Herrero}, A., \&
  {P{\'e}rez-Gonz{\'a}lez}, P.~G. 2006, \apj, 645, 115

\bibitem[{{Risaliti} {et~al.}(2005){Risaliti}, {Elvis}, {Fabbiano}, {Baldi}, \&
  {Zezas}}]{Risaliti05}
{Risaliti}, G., {Elvis}, M., {Fabbiano}, G., {Baldi}, A., \& {Zezas}, A. 2005,
  \apjl, 623, L93

\bibitem[{{Risaliti} {et~al.}(1999){Risaliti}, {Maiolino}, \&
  {Salvati}}]{Risaliti99}
{Risaliti}, G., {Maiolino}, R., \& {Salvati}, M. 1999, \apj, 522, 157

\bibitem[{{Rosen} {et~al.}(2016){Rosen}, {Webb}, {Watson}, {Ballet}, {Barret},
  {Braito}, {Carrera}, {Ceballos}, {Coriat}, {Della Ceca}, {Denkinson},
  {Esquej}, {Farrell}, {Freyberg}, {Gris{\'e}}, {Guillout}, {Heil},
  {Koliopanos}, {Law-Green}, {Lamer}, {Lin}, {Martino}, {Michel}, {Motch},
  {Nebot Gomez-Moran}, {Page}, {Page}, {Page}, {Pakull}, {Pye}, {Read},
  {Rodriguez}, {Sakano}, {Saxton}, {Schwope}, {Scott}, {Sturm}, {Traulsen},
  {Yershov}, \& {Zolotukhin}}]{Rosen16}
{Rosen}, S.~R., {Webb}, N.~A., {Watson}, M.~G., {et~al.} 2016, \aap, 590, A1

\bibitem[{{Rovilos} {et~al.}(2011){Rovilos}, {Fotopoulou}, {Salvato},
  {Burwitz}, {Egami}, {Hasinger}, \& {Szokoly}}]{Rovilos11}
{Rovilos}, E., {Fotopoulou}, S., {Salvato}, M., {et~al.} 2011, \aap, 529, A135

\bibitem[{{Rovilos} {et~al.}(2014){Rovilos}, {Georgantopoulos}, {Akylas},
  {Aird}, {Alexander}, {Comastri}, {Del Moro}, {Gandhi}, {Georgakakis},
  {Harrison}, \& {Mullaney}}]{Rovilos14}
{Rovilos}, E., {Georgantopoulos}, I., {Akylas}, A., {et~al.} 2014, \mnras, 438,
  494

\bibitem[{{Schawinski} {et~al.}(2007){Schawinski}, {Thomas}, {Sarzi},
  {Maraston}, {Kaviraj}, {Joo}, {Yi}, \& {Silk}}]{Schawinski07}
{Schawinski}, K., {Thomas}, D., {Sarzi}, M., {et~al.} 2007, \mnras, 382, 1415

\bibitem[{{Shi} {et~al.}(2010){Shi}, {Rieke}, {Smith}, {Rigby}, {Hines},
  {Donley}, {Schmidt}, \& {Diamond-Stanic}}]{Shi10}
{Shi}, Y., {Rieke}, G.~H., {Smith}, P., {et~al.} 2010, \apj, 714, 115

\bibitem[{{Stern} \& {Laor}(2012)}]{SternLaor12}
{Stern}, J. \& {Laor}, A. 2012, \mnras, 423, 600

\bibitem[{{Szokoly} {et~al.}(2004){Szokoly}, {Bergeron}, {Hasinger}, {Lehmann},
  {Kewley}, {Mainieri}, {Nonino}, {Rosati}, {Giacconi}, {Gilli}, {Gilmozzi},
  {Norman}, {Romaniello}, {Schreier}, {Tozzi}, {Wang}, {Zheng}, \&
  {Zirm}}]{Szokoly04}
{Szokoly}, G.~P., {Bergeron}, J., {Hasinger}, G., {et~al.} 2004, \apjs, 155,
  271

\bibitem[{{Tran}(2001)}]{Tran01}
{Tran}, H.~D. 2001, \apjl, 554, L19

\bibitem[{{Tran}(2003)}]{Tran03}
{Tran}, H.~D. 2003, \apj, 583, 632

\bibitem[{{Tran} {et~al.}(2011){Tran}, {Lyke}, \& {Mader}}]{Tran11}
{Tran}, H.~D., {Lyke}, J.~E., \& {Mader}, J.~A. 2011, \apjl, 726, L21

\bibitem[{{Tran} {et~al.}(1992){Tran}, {Osterbrock}, \& {Martel}}]{Tran92}
{Tran}, H.~D., {Osterbrock}, D.~E., \& {Martel}, A. 1992, \aj, 104, 2072

\bibitem[{{Tremonti} {et~al.}(2004){Tremonti}, {Heckman}, {Kauffmann},
  {Brinchmann}, {Charlot}, {White}, {Seibert}, {Peng}, {Schlegel}, {Uomoto},
  {Fukugita}, \& {Brinkmann}}]{Tremonti04}
{Tremonti}, C.~A., {Heckman}, T.~M., {Kauffmann}, G., {et~al.} 2004, \apj, 613,
  898

\bibitem[{{Urry} \& {Padovani}(1995)}]{Urry95}
{Urry}, C.~M. \& {Padovani}, P. 1995, \pasp, 107, 803

\bibitem[{{Veilleux} {et~al.}(1997){Veilleux}, {Goodrich}, \&
  {Hill}}]{Veilleux97}
{Veilleux}, S., {Goodrich}, R.~W., \& {Hill}, G.~J. 1997, \apj, 477, 631

\bibitem[{{Wang} {et~al.}(2004){Wang}, {Malhotra}, {Rhoads}, \&
  {Norman}}]{Wang04}
{Wang}, J.~X., {Malhotra}, S., {Rhoads}, J.~E., \& {Norman}, C.~A. 2004, \apjl,
  612, L109

\bibitem[{{Ward} {et~al.}(1987){Ward}, {Geballe}, {Smith}, {Wade}, \&
  {Williams}}]{Ward87}
{Ward}, M.~J., {Geballe}, T., {Smith}, M., {Wade}, R., \& {Williams}, P. 1987,
  \apj, 316, 138

\bibitem[{{Xiao} {et~al.}(2011){Xiao}, {Barth}, {Greene}, {Ho}, {Bentz},
  {Ludwig}, \& {Jiang}}]{Xiao11}
{Xiao}, T., {Barth}, A.~J., {Greene}, J.~E., {et~al.} 2011, \apj, 739, 28

\end{thebibliography}

\end{document}